\documentstyle[12pt]{article}
\begin{document}
\rightline{DOE/ER/40427-28-N94}
\rightline{November 29, 1994 }
\vspace{9mm}

\begin{center}
{\bf COLOR  TRANSPARENCY  EFFECTS   IN \\
ELECTRON DEUTERON INTERACTIONS AT INTERMEDIATE {\boldmath $Q^2$}}

\vspace{2mm}

L.~L.~Frankfurt\\
{\it School of Physics and Astronomy, Tel Aviv University, 69978, Israel\\
Institute for Nuclear Physics, St. Petersburg, Russia}

\vspace{1mm}

W.~R.~Greenberg\\
{\it Department of Physics, University of Pennsylvania,
Philadelphia, PA 19104}

\vspace{1mm}

G.~A.~Miller\\
{\it Department of Physics, University of Washington, Seattle, WA 98195}

\vspace{1mm}

M.~M.~Sargsyan\\
{\it School of Physics and Astronomy, Tel Aviv University,  69978, Israel, \\
Yerevan Physics Institute, 375036, Armenia}
\vspace{1mm}

and

\vspace{1mm}

M.~I.~Strikman\\
{\it Department of Physics, Pennsylvania State University,
University Park, PA 16802\\
Institute for Nuclear Physics, St. Petersburg, Russia}
\end{center}

\vspace{1cm}
\begin{abstract}
High momentum transfer electrodisintegration of polarized and unpolarized
deuterium targets, $d(e,e'p)n$ is studied. We show that the importance of
final state interactions FSI, occuring when a knocked out nucleon interacts
with the other nucleon, depends strongly on the momentum $\vec p_n$ of the
spectator nucleon.  In particular, these FSI  occur when the essential
contributions to the scattering amplitude arise from internucleon distances
$\sim 1.5~fm$. But the absorption of the high momentum $\gamma^*$ may produce
a point like configuration, which evolves with time. In this case, the final
state interactions  probe the point like configuration at the early stage of
its evolution. If the point like configuration is still small after propagating
about 1.5 fm, the FSI are suppressed. The result is that significant color
transparency effects, which can either enhance or suppress computed cross
sections,  are predicted to occur for $\sim 4 GeV^2 \ge Q^2\leq~10~(GeV/c)^2$.
We suggest searching  for color transparency phenomenon by examining ratios of
experimentally measured quantities. Possible theoretical uncertainties of the
calculations,including those due to the deuteron wave function and relativistic
effects, are found to be small.
\end{abstract}


\section{Introduction}

    Color transparency (CT) and color coherent effects have
been recently under
intense experimental and theoretical investigation. The (p,pp) experiment of
Carroll et al.\cite{Carol1} found indications of color transparency while
the NE18 (e,e'p) experiment \cite{NE18} found no such indications. The
appearance of color transparency depends on formation of a point-like
configuration (PLC) by hard scattering. The  Q$^2$ of the NE18 experiment
(1 $\le Q^2 \le 7 GeV^2$) seem to be  large enough to form a small color
singlet object. The reason why significant color transparency is not observed
is the rapid expansion of PLC to nearly normal size (a nearly normal
 absorption)
at the relatively low momenta of the ejected protons\cite{FLFS,GM}.
 Thus models
of color transparency which reproduce the (p,2p) data and include expansion
effects predicted small CT effects for the NE-18 kinematics, consistent with
their findings, see the discussion in Ref.\cite{ANN}.

Therefore a new strategy  of using light nuclei and double scattering as a
means to investigate whether small objects are produced at intermediate $Q^2$
was suggested in Ref.\cite{EFGMSS94}. The idea was that using the lightest
nuclei would allow the suppression of expansion effects, while employing
kinematical cuts to suppress the plane wave Born contributions and to enhance
the contribution of double scattering interactions would increase the
deviations
from the Glauber model. That work concerned  $He$  nuclei target, in the
 present
we use an improved formalism to investigate the use of a deuteron target.

     The deuteron is the best understood nuclear system, with  a wave
function  determined experimentally in a wide momentum range \cite{Brown}.
Therefore finding a method to use the deuteron to investigate color coherent
-color transparency effects is both challenging and potentially rewarding.
The average separation between the nucleons in the deuteron is large, so the
deviations from the plane wave impulse approximation (PWIA) computed within
the ordinary Glauber approximation are usually very small, much smaller than
for a $He$ target.

However a recent analysis of hard quasielastic $d(p,2p)$ reactions within the
framework of conventional Glauber approximation \cite{ELI} has found
substantial Glauber screening effects for transverse spectator momenta less
than $300~MeV/c$ and small longitudinal spectator momentum. In particular,
the interference between the Born term  and rescattering amplitude in which
the knocked out nucleon interacts with the spectator nucleon, causes a
reduction
in the cross section. This is called a screening effect. Including the
 square of
the rescattering amplitude, which we call the double scattering term,
increases
the computed cross section. Color transparency  leads to a suppression of the
rescattering amplitude, predicting specific upward and downward changes in the
cross section, depending on the kinematics which control the relative sizes of
the screening and double  scattering terms. Thus observing such effects which
occur for different values of the spectator momenta seems to be an effective
method to search for color coherent effects using the deuteron.

Substantial screening has been observed  also in the recent calculation
\cite{NIC} of the processes $d(e,e'p)$  within the nonrelativistic Glauber
approximation. The present paper uses formulae of the
Glauber approximation deduced from the  Feynman diagrams.
 As a result the predicted angular dependence of the cross
section at higher spectator momenta is  noticeably different from that in
Ref.\cite{NIC}. In particular, the  position in
the  spectator angular distribution,
corresponding to the  maximal final state interaction
 depends on the momentum of the
spectator nucleon. At $Q^2\approx 1(GeV/c)^2$ our calculations agree well with
calculations of Ref.\cite{AREN0,LEID} in the kinematics
where the double scattering
term dominates. According to Ref.\cite{AREN0} in this
 kinematics the contribution
of meson exchange currents and isobars in the intermediate
 is small. This is another example that the
Glauber approximation is a good starting point in the hunt for the
 CT phenomenon.

We stress that  exclusive  processes of high momentum transfer
electrodisintegration of the polarized and unpolarized deuteron -
$d(e,e'p)n$ have an important advantage. One may  choose  special kinematical
conditions for the spectator nucleon to control the essential internucleon
distances ($\sim 1.5~fm$) for final state interactions (FSI). Therefore it is
possible to probe the PLC at the early stage
of its evolution, where expansion effects are minimal. The chosen kinematics
also allow the separate investigation  of the different implications of CT for
the
screening and double scattering terms. Another limitation on the kinematics is
the requirement of reliability of the nonrelativistic description of deuteron
 and
smallness of competing nuclear effects which might diminish  CT effects. The
major
price for fixing kinematics is the drop in the cross section, which should not
cause a problem for high intensity accelerators.

The paper is organized as follows. In section 2 we calculate the cross
section of
the $d(e,e'p)n$ reaction within the framework of the conventional Glauber
approximation\cite{GA}. Polarized and unpolarized targets are studied.

In section 3  color transparency effects in the $d(e,e'p)n$ reaction are
calculated within the framework of quantum diffusion model\cite{FLFS} and
three resonance model\cite{FGMS93}. The three resonance model is solved in
two ways: the Green function method of Ref.\cite{EFGMSS94} and the modified
Glauber approximation method, used previously to calculate  nuclear vector
meson production (see  Refs.\cite{YN,KM,BAU70,BAR}).

Section 4 contains the numerical results. We first use the Glauber
approximation to investigate the kinematic requirements necessary to ensure
the most favorable conditions for studying color coherent effects. This
analysis allows us to introduce new experimental observables which are free
from theoretical normalizations and are more sensitive to the effects of final
state interactions. Using the models, discussed in section 3, we calculate
 color
transparency effects for the kinematic range accessible to the present
generation
of electron accelerators.

Section 5 considers theoretical uncertainties and the reliability of obtained
results.

The main results of present paper are summarized in Section 6.

\section{Glauber Approximation (GA)}

    We consider the $d(e,e'p)n$ reaction for kinematics in which the proton
carries almost all of the momentum of  of the virtual photon and  the neutron
momentum
$p_n$ is in the range  200-400~$MeV/c$ where deuteron wave functions have been
determined experimentally.
\footnote{Obviously, kinematics in which the neutron is produced forward and
proton sidewise is as good for our purposes. In fact, such a setup may
have certain experimental advantages since in this case one detects
neutron with the same momentum for different $Q^2$. We thank B.Mecking
for emphasizing this point.}
The amplitude for $d(e,e'p)n$ processes can be expressed as the matrix element
of the scattering operator - $\hat{T}$ between the deuteron wave  function and
the wave function of the final state consisting  of two  outgoing  nucleons:
\begin{equation}
{\cal M} = \int\int \langle \vec{p}_{p}, s_{p}; \vec{p}_{n}; s_{n} \mid
\hat{T} (q) \mid d,\vec{s}  \rangle d^3r_p d^3r_n.
\label{GN}
\end{equation}
Here $\mid d,\vec{s} \rangle$ is  the position space deuteron wave function,
$\vec p_p$, $\vec r_p$, $s_p$ and $\vec p_n$, $\vec r_n$, $s_n$ are the
momenta,  coordinate and spin of proton and neutron in the final state.

   Within the conventional assumptions on factorization of hard -
electromagnetic
and soft - FSI amplitudes i.e. within the Glauber approximation the  nuclear
scattering operator has the coordinate space form:
\begin{equation}
\hat{T}(q,r_p,r_n) =
T_{S}(p_p,p_n,r_p,r_n)\cdot  T_H^{em}(Q^2)\cdot  e^{i\vec q \cdot \vec r_p},
\label{OP}
\end{equation}
where  $T_{H}^{em}(Q^2)$ - is  the one-body electromagnetic  current operator,
$\vec q$ is the space component of the photon four-momentum which we choose in
the z-direction.  We note that factorization is violated when off-shell
 effects
are taken into account. So we will focus on CT effects at small spectator
 nucleon
momenta or in the rescattering region where only  small  momenta in the loop
integral over internal nucleon momenta are important. We neglect the
contributions
of meson currents like that at Fig.1c since such contributions rapidly
decrease
with  $Q^2$ for $x\sim 1$ (cf. discussion in  \cite{FGKS89} and sec. 5.5
bellow).
$T_{S}(p_p,p_n)$  describes soft FSI of knocked-out proton with spectator
 neutron.
We are studying situations in which the momentum of knocked - out proton
 is larger
than $1~GeV/c$. So  $T_{S}$  can  be calculated within the Glauber
approximation
\cite{GA,FG,BERT}
\footnote{To simplify the discussion we neglect small corrections
due to the charge exchange process $np \rightarrow pn$.
Since the charge exchange amplitude is predominantly real and of spin flip
character
(which itself is small), it practically does not interfere with the main
amplitude.}:
\begin{equation}
T_{S}(p_p,p_n,r_p,r_n)  = 1-\Gamma^N(b_p-b_n)\cdot\Theta(z_n-z_p)\cdot
e^{-i\Delta^0(z_p-z_n)},
\label{GLS}
\end{equation}
where $\vec r~\equiv~\vec r(z,\vec b)$. Here the $z$ direction is defined
 by the
direction of the struck proton momentum $\vec p_p$. However within the
considered
kinematics, where proton carries almost all momentum of virtual photon,
 we choose
for $z$ the direction of $\vec q$. The profile function $\Gamma^N(b)$
is expressed
via the nucleon -nucleon scattering amplitude ($f^{NN}$) as follows:
\begin{equation}
\Gamma^N(b) =
{1\over 2 i }\int \exp{(i{\vec k_t \cdot \vec b})} \cdot f^{NN}({\vec  k_t})
{d^2k_t\over  (2\pi)^2}  \label{gamma},
\end{equation}
where the $NN$ scattering amplitude normalized as
$Imf^{NN}(k_t=0)~=~\sigma_{tot}$.
The additional factor $e^{-i\Delta^0(z_p-z_n)}$ in eq.(\ref{GLS}), where
\begin{equation}
\Delta^0 = (E_n-m)\cdot {m_d+q_0\over |\vec q|},
\label{RF}
\end{equation}
is a kinematical factor. In eq.(\ref{RF}) $E_n$ - is the spectator
 energy, $m$ is the nucleon mass,
$m_d$ - is
the mass of deuteron and $q_0$ is the transfered energy. The factor
 $\Delta^0$  is obtained by evaluating
Feynman diagrams (cf. section 5.4, eq.(\ref{lcom}))
and accounts for the nucleon recoil
usually neglected within the
conventional non-relativistic Glauber approximation. This factor accounts
 for the fact
that the variable ${k_0-k_z\over m}$ but not $k_z$ (fig.1b) is conserved in
 two-body high
energy collisions (cf. \cite{FS81}).

Inserting eqs.(\ref{GLS}) and (\ref{gamma}) into  eq.(\ref{GN}), we obtain
the amplitude for the  $d(e,e'p)n$ reaction as:
\begin{eqnarray}
{\cal M}  &  = &T_H^{em}(Q) \int\int d^3r_pd^3r_n  e^{i\vec q \cdot\vec r_p}
\langle \vec{r}_p,s_p;\vec{r}_{n}, s_{n} \mid d, \vec{s} \rangle
\nonumber \\
& & \times \left [1 -  \Gamma(b_p-b_n)e^{-i\Delta^0(z_p-z_n)}
\Theta(z_n-z_p)\right]
e^{-i\vec p_p  \cdot\vec r_p} e^{-i\vec p_n \cdot \vec r_n} \nonumber \\
&  = &   T_H^{em}(Q)\int\int d^3r_pd^3r_n
e^{-i\vec p_i \cdot\vec r_p} e^{-i\vec p_n \cdot\vec r_n}
\langle \vec{r}_{p}, s_{p}; \vec{r}_{n}, s_{n} \mid d,\vec{s}\rangle\nonumber
\\
& &  -  T_H^{em}(Q) \int\int\limits_{z_p} d^3r_pd^3r_n
e^{-i\vec p_i \cdot\vec r_p} e^{-i\vec p_n \cdot\vec r_n}
\langle \vec{r}_{p}, s_{p}; \cdot \vec{r}_{n},s_{n} \mid d,\vec{s} \rangle
 \nonumber \\
& & \times {1\over 2 i }\int e^{i{\vec k_t \cdot(\vec b_p-\vec b_n })}
\cdot f^{NN}({\vec  k_t}){d^2k_t\over  (2\pi)^2}
e^{-i\Delta^0(z_p-z_n)}\Theta(z_n -z_p),
\label{GA_M2}
\end{eqnarray}
where we define the momentum of initial proton in the
deuteron state as  $\vec p_i = \vec p_p - \vec q$.

To separate center-of-mass motion we introduce the center-of-mass
 ($R_{cm}$) and
relative distance ($r$) coordinates:
\begin{eqnarray}
\vec r_p & = & \vec R_{cm} + {1\over  2}\vec r \nonumber \\
\vec r_n & = & \vec R_{cm} - {1\over 2}\vec r.
\label{REL}
\end{eqnarray}
Using Eq.(\ref{REL}) allows us to simplify Eq.(\ref{GA_M2}) as:
\begin{eqnarray}
{\cal M} & = & T_p^{em}(Q) \int d^3R_{cm}e^{i\vec R_{cm}\cdot
(\vec p_{d} - \vec p_i - \vec p_n)}
\left\{ \int d^3r \langle \vec{r}_{p},s_{p}; \vec{r}_{n},s_{n} \mid d,
\vec{s} \rangle
e^{-{i\over 2}\vec p_i \cdot \vec r} e^{{i\over 2}\vec p_n \cdot \vec r}
\right.
\nonumber \\
& - & \left. \int d^3r \Theta(-z) \langle
\vec{r}_{p},s_{p};\vec{r}_{n},s_{n} \mid d, \vec{s} \rangle\cdot
e^{-{i\over 2}\vec p_i \cdot \vec r} e^{{i\over 2}\vec p_n \vec r}
{1\over 2 i }\int e^{i\vec k_t \cdot \vec b -i\Delta^0z}
\cdot f^{NN}({\vec  k_t}){d^2k_t\over  (2\pi)^2}\right\}\nonumber \\
& = & T_p^{em}(Q) \left\{\int d^3r \langle \vec{r}_{p}, s_{p};
\vec{r}_{n}, s_{n}\mid d, \vec{s} \rangle e^{-i\vec p_i \vec r}
\right.\nonumber \\
& - &\left.  {1\over 2i} \int\int d^3r \Theta(-z)
\langle \vec{r}_{p}, s_{p}; \vec{r}_{n}, s_{n} \mid d,\vec{s}\rangle\cdot
e^{-i(\vec p_{it} - \vec k_t)\cdot \vec b - i(p_{iz}+\Delta^0)\cdot z}
 f^{NN}({\vec  k_t}){d^2k_t\over  (2\pi)^2}\right\},
\label{GA_M3}
\end{eqnarray}
The integration over $R_{cm}$ ensures the momentum conservation:
$\vec p_n = -\vec p_i = \vec q - \vec p_p$. In the last part of
eq.(\ref{GA_M3}) we omit the factor
$(2\pi)^3 \delta(\vec p_d - \vec p_p - \vec p_n)$,
which will be included  in the definition of cross section.
(To simplify formulae we use the Laboratory  frame, in which the momentum of
of the  deuteron, $\vec p_d$ is zero.)

The integration over the relative coordinate $r$ leads to the result:
\begin{equation}
{\cal M} =  (2\pi)^{{3\over 2}}\cdot
T_p^{em}(Q)\left\{  \langle p_{i},s_{p},s_{n}\mid d,\vec{s}\rangle
- {1\over 4i} \int \langle p_{i}^{'}, s_{p}, s_{n}\mid d,\vec{s}\rangle
f^{NN}({\vec  k_t}){d^2k_t\over  (2\pi)^2}\right\},
\label{GA_M4}
\end{equation}
where $\vec{p_i'}~\equiv~(p_{iz}+\Delta^0,p_{it}-k_t)$ and
$\langle p_{i}^{'}, s_{p}, s_{n} \mid d, \vec{s} \rangle$ - is the deuteron
wave function in momentum space representation. The  additional factor
${1\over 2}$ in the  second term is a consequence of $\Theta(-z)$ functions
in eq.(\ref{GA_M3}). (The accuracy of last replacement
 is discussed in sec.5.4).

Taking the square of the modulus of the amplitude and summing over the
proton and neutron polarizations we obtain:
\begin{eqnarray}
\mid\mid{\cal M}\mid\mid ^2 & \equiv & \sum_{s_p,s_n,s_p',s_n'}{\cal M}{\cal M}
^{\dag}
\nonumber \\
& = &  (2\pi)^3\cdot
|F^{e.m.}_p (Q^2)|^2\cdot \left\{\rho^{\vec s} _d(p_i,p_i) - {\cal R}e
{1\over 2i}
\int {d^2k_t\over (2\pi)^2} \vec{\rho^s_d}(p_i',p_i)f^{NN*}(k_t) \right.
\nonumber  \\
& & +  \left. {1\over 16}\int\int {d^2k_{t1}d^2k_{t2}\over (2\pi)^4}
\vec{\rho^s_d}(p_{i1},p_{i2})f^{NN}(k_{t1}) f^{NN*}(k_{t2}) \right\},
\label{GA2}
\end{eqnarray}
where
$\vec{p_{i1}}~\equiv~(p_{iz}+\Delta^0,p_{it}-k_{t1})$,
$\vec{p_{i2}}~\equiv~(p_{iz}+\Delta^0,p_{it}-k_{t2})$.

In  eq.(\ref{GA2}) we introduced the deuteron density  function:
\begin{equation}
\rho^{\vec s}(k_1,k_2) = \sum_{s_p,s_p',s_n,s_n'}
\langle k_{2},s_{p}, s_{n} \mid d,\vec{s} \rangle \langle
d, \vec{s} \mid k_{1},s_{p}^{'}, s_{n}^{'} \rangle .
\label{RHO}
\end{equation}

In this Glauber or distorted  wave impulse
approximation (DWIA) the $d(e,e'p)n$
cross section  can be expressed as follows:
\begin{equation}
{d\sigma\over dE_{e'} d\Omega_{e'} d^3p_p } =
\sigma_{ep}\cdot D_{d}(q,p_p,p_n)\cdot
\delta(q_o - M_d - E_p - E_n).
\label{CRS}
\end{equation}
Here  $\sigma_{en}$  is the cross section of the electron  scattering off a
bound proton (up to the flux and proton recoil factor)
\footnote{In principle, one should use   the light-cone quantum mechanics of
the deuteron \cite{FS81} to calculate the cross section.
However for the effects dominated by the contribution of small nucleon momenta
in the deuteron of interest here, the difference between predictions of
light-cone and
nonrelativistic formalisms is small
(see Sec.5.1).} and the decay function $D_{d}(q,p_p,p_n)$ represents the joint
probability for the initial proton in the deuteron  having Fermi momentum -
$p_i$ and for the final state having a proton and neutron with a momentum
$p_p$ and $p_n$ (for more details see \cite{FS88}).

If the deuteron is polarized, the cross section depends on
$D_{d}^{\vec {s}}(q,p_p,p_n)$ with:
\begin{eqnarray}
D_{d}^{\vec {s}}(q,p_p,p_n) & = & \rho^{\vec s} _d(p_i,p_i) -
{\cal R}e{1\over 2i} \int \rho^{\vec s} _d(p_i,p_i')\cdot f^{NN}(k_t)\cdot
{d^2k_t\over (2\pi)^2} \nonumber \\
& & + {1\over 16} \int \rho^{\vec s} _d(p_{i1},p_{i2})\cdot
f^{NN}(k_{t1})\cdot
f^{NN*}(k_{t2})\cdot
{d^2k_{t1}\over (2\pi)^2} {d^2k_{t2}\over (2\pi)^2}.
\label{D_GA}
\end{eqnarray}
The polarized density matrices of deuteron can be  expressed via the
$s$ and $d$ - wave components introducing the polarization vector
$\vec a$ according to Ref.\cite{AB}:
\begin{eqnarray}
\rho_{d}^{\vec a}(k_1,k_2)    =   u(k_1)u(k_2) +
\left[1 - {3|k_2\cdot a|^2 \over k_2^2} \right]{u(k_1)w(k_2) \over \sqrt{2}} +
\left[1 - {3|k_1\cdot a|^2 \over k_1^2} \right]{u(k_2)w(k_1) \over \sqrt{2}}
\nonumber \\
\nonumber \\
+\left( {9\over 2} { (k_1 \cdot a)(k_2\cdot a)^*(k_1\cdot k_2)
 \over k_1^2k_2^2 }
 -
{3\over2}{|k_1\cdot a|^2\over k_1^2} - {3\over2}{|k_2\cdot a|^2\over k_2^2} +
{1\over 2}
\right) w(k_1)w(k_2),
\nonumber \\
\label{RHO_A}
\end{eqnarray}
where the components of $\vec a$ are defined through the deuteron spin wave
function as:
\begin{equation}
\psi^{10} = i\cdot a_z , \  \psi^{11} = -{i\over \sqrt{2}}(a_x + ia_y) ,
                           \  \psi^{1-1} = {i\over \sqrt{2}}(a_x -
ia_y),
\label{VK}
\end{equation}
where $\psi^{1\mu}$ is the projection of the deuteron's spin on the the $\mu$
direction. The unpolarized deuteron density matrix follows from
Eq.(\ref{RHO_A})
as:
\begin{equation}
\rho_d(k_1,k_2)  =  {1\over 3} \sum_a  \rho_{d}^{a}(k_1,k_2) =
 u(k_1)u(k_2) + w(k_1)w(k_2)\cdot
\left({3\over 2}{(\vec{k}_1\cdot \vec{k}_2)^2\over k_1^2k_2^2} -
{1\over 2}\right).
\label{RHO_UP}
\end{equation}

\section{Coherent Effects in  {\boldmath $d(e,e'p)n$}   Scattering}

Theoretical analysis shows that in  realistic models the absorption of a hard
photon leads to the formation of a point like configuration, which undergoes a
reduced interaction with other hadrons, because of its small size and its
 color
neutrality \cite{FMS92}. To estimate the expected effects of color coherence
we consider two different models which account for the formation of the
 PLC and
their evolution to the normal hadronic state: quantum diffusion model of
Ref.\cite{FLFS} and the three state model of Ref.\cite{FGMS93}.

\subsection{Quantum Diffusion model (QDM)}

The reduced interaction  between the PLC and the spectator neutron can be
accounted for by introducing the dependence of the  scattering amplitude on
the transverse size of PLC. However we  consider  energies that are far from
asymptotic, so  the expansion of PLC  should be important. This feature is
included by allowing the rescattering amplitude to depend on the distance from
the photon absorption point. Including these effects leads to the modified
deuteron decay function  $D_d^S(q,p_p,p_n)$:
\begin{eqnarray}
D_{d}^{\vec s}(q,p_p,p_n)  & = &  \rho^s_d(u(p_i),w(p_i),u(p_i),w(p_i))
\nonumber \\
& & - {\cal R}e{1\over i}\int \rho^{\vec s}_d(u(p_i),w(p_i),\tilde
u(p_i')^*,\tilde w(p_i')^*) f^{NN}(k_t)
{d^2k_t\over (2\pi)^2} \nonumber \\
& & + {1\over 4} \int
\rho^s_d(\tilde u(p_{i1}), \tilde w(p_{i2}),\tilde u(p_i'),\tilde w(p_i'))
{d^2k_{t1}\over (2\pi)^2}f^{NN}(k_{t1}) f^{NN*}(k_{t2}) {d^2k_{t2}\over
(2\pi)^2},
\nonumber \\
\label{D_DIF}
\end{eqnarray}
with the same form of $\rho^{\vec s}_d$ as in Eq.(\ref{RHO_A}) but
with modified radial wave functions:
\begin{eqnarray}
\tilde u(k) & = & {1\over (2\pi)^{{3\over 2}}}
\int u(r)Y_o(\hat r)\cdot {f^{PLC,N}(z,k_t,Q^2) \over f^{NN}(k_{t})}
 \Theta (-z)
e^{-i\vec k \cdot \vec r} d^3r
\nonumber \\
\tilde w(k) & = & {1\over (2\pi)^{{3\over 2}}}
\int w(r)Y_{20}(\hat r)\cdot  {f^{PLC,N}(z,k_t,Q^2) \over f^{NN}(k_{t})}
\Theta (-z) e^{-i\vec k \cdot \vec r} d^3r.
\label{TUW}
\end{eqnarray}

Eq.(\ref{D_DIF}) is derived in the same way as the  Glauber formulae in the
previous section.  The only difference is to include the  coordinate
 dependence
of the rescattering amplitude. To calculate the deuteron decay function in the
quantum diffusion model we model the amplitude of the $PLC-N$ scattering in
a form\cite{EFGMSS94} consistent with the optical theorem:
\begin{equation}
f^{PLC,N}(z,k_t,Q^2) = i\sigma_{tot}(z,Q^{2}) \cdot
e^{{b\over 2 }t}\cdot {G_{N}(t\cdot\sigma_{tot}(z,Q^{2})/\sigma_{tot})
\over G_{N}(t)},
\label{F_NNCT}
\end{equation}
where $b/2$ is the slope of elastic $NN$ amplitude,    $G_{N}(t)$
($\approx  (1-t/0.71)^{2}$) is the Sachs form factor and
$t= -k_t^2 $. The last factor in eq.(\ref{F_NNCT}) accounts for
the difference  between elastic  scattering of  PLC and average
configurations, using  the observation that the $t$ dependence
of $d\sigma^{h+N\rightarrow h+N}/dt $ is roughly that of
$\sim~G_{h}^{2}(t)\cdot G_{N}^{2}(t)$ for not very large values of t
and that $G_{h}^{2}(t)\approx exp(R_h^2t/3)$.

In Eq.~(\ref{F_NNCT}) $\sigma_{tot}(l,Q^{2})$  is the  effective total
cross section of the  interaction  of the PLC at the distance $l$ from
the interaction point. The quantum diffusion   model~\cite{FLFS}
corresponds to:
\begin{equation}
\sigma _{tot}(l,Q^{2}) = \sigma_{tot} \left \{ \left ({l \over l_{h}} +
{\langle r_{t}(Q^2)^{2} \rangle \over \langle r_t^{2}  \rangle }
(1-{l \over l_{h}}) \right )\Theta (l_{h}-l) + \Theta (l-l_{h})\right\},
\label{SIGMA_CT}
\end{equation}
where ${l_h = 2p_{f}/\Delta~M^{2}}$, with ${\Delta~M^{2}=0.7-1.1~GeV^{2}}$.
Here ${\langle r_{t}(Q^2)^{2} \rangle}$  is the average transverse size
squared of the  configuration  produced at the interaction  point.
In several realistic models considered  in Ref.\cite{FMS92} it can be
approximated as ${ {\langle r_{t}(Q^2)^2\rangle\over\langle r_t^2\rangle}
\sim{1\,GeV^2\over Q^2}}$ for  $Q^2~\geq~1.5~GeV^2$.  Note that due to
effects of expansion the results of calculations are rather insensitive
to the value of this ratio whenever it is much less than unity.

\subsection {Three-State Model}

\subsubsection {Summary of the model}

 To evaluate matrix element of operators $T_S$ and $T_H$ in eq.(\ref{GN})
we use the three-state model of Ref.\cite{FGMS93}. The basic assumption
of this model is that the hard scattering operator $T_H$ acts on a nucleon to
produce a non-interacting point-like configuration $|PLC\rangle$ which is a
superposition of three baryonic states:
\begin{equation}
T_H |N\rangle = |PLC\rangle = \sum_{m=N,N^*,N^{**}} F_{m,N}(Q^2) | m \rangle,
\label{TH}
\end{equation}
where $F_{m,N}(Q^2)$  are elastic ($m=N$) and inelastic
 transition form factors
in this model. It is important to  notice that  the transition
 form factors used
here cannot be taken directly from data. This is because states
 $N^*$ and $N^{**}$
are {\it effective} states which, in fact, represent a number of
 actual physical
states. In the following analysis we assume for certainty that all
form factors
have the same $Q^2-$dependence. We neglect also possible spin effects in the
form factors and in the operator $T_H$.

Color transparency is introduced in this model as the condition of lack of FSI
at the point where PLC is produced:
\begin{equation}
T_S|PLC\rangle = 0.
\label{SUM}
\end{equation}
It is this condition  which distinguishes models with color transparency
from models which include production of  resonances and/or continuum
in intermediate states\cite{KOP}. Within the approximations discussed
above $T_S$ is equal to a matrix  $U$ which is the most general
3 $\times$ 3 (Hermitian) matrix that annihilates the PLC:
\begin{equation}
U= \sigma^{tot} \left( \begin{array}{ccc}
1 & { {-F_{N,N} + \epsilon F_{N^{**},N} }\over F_{N^*,N}} & - \epsilon \\
\cdots & \mu &
{ {|F_{N,N}|^2 - \epsilon^* F_{N,N}F_{N^{**},N}^*-\mu |F_{N^*,N}|^2}
\over{F_{N^*,N}^*F_{N^{**},N}}} \\
\cdots & \cdots &
{ {\mu |F_{N^*,N}|^2 - |F_{N,N}|^2 + 2 {\rm Re}(\epsilon^*F_{N,N}
F_{N^{**},N}^*)}
\over {|F_{N^{**},N}|^2}}
\end{array} \right).
\label{UM}
\end{equation}
Taking account of the data on  $pp$ and $pd$ diffractive scattering helps to
restrict further  the parameters of this matrix\cite{FGMS93}. We use the
matrix $U$ from that work.

\subsubsection{Green Function Method}
In this section we use the Green function method developed in
Ref.\cite{EFGMSS94}
to calculate CT effects in the double scattering reaction off
 $^3He$. Within this
method, the matrix element for the process $d(e,e'p)n$ which
 includes final-state
interactions, is equal to
\begin{equation}
{\cal M}^s(\vec p_p, s_p; \vec p_n, s_n) = \langle \vec p_p, s_p;
\vec p_n, s_n | T_H + T_S G_0 T_H | d, \vec s \rangle,
\label{eq:scattering}
\end{equation}
where $\vec p_{p(n)}$ is the momentum of the recoil proton (neutron),
$s_{p(n)}$ is the projection of the recoil proton (neutron) spin and
$s$ is the projection of the deuteron spin. $T_H$ is the hard scattering
operator - the electromagnetic current. $G_0$ is the free propagator, and
$T_S$ accounts for the soft final-state interaction. The matrix element
${\cal M}^s$ is also a function of the four-momentum transfer $Q^2$ and
the Bjorken variable $x$ (or equivalently the virtual photon three-momentum
$\vec q$ and its energy $q^0$). However, this dependence will be omitted
in the formulae to simplify notations. It is convenient to define
\begin{eqnarray}
{\cal M}^s_0(\vec p_p, s_p; \vec p_n, s_n) &=& \langle \vec p_p, s_p;
\vec p_n, s_n | T_H  | d, \vec s \rangle,\\
{\cal M}^s_1(\vec p_p, s_p; \vec p_n, s_n) &=& \langle \vec p_p, s_p;
\vec p_n, s_n | T_S G_0 T_H | d, \vec s \rangle.
\end{eqnarray}
The matrix element for the hard scattering operator is given by
(cf  eq.(\ref{TH}))
\begin{equation}
\langle m, {\vec r}_p, s_p; n, {\vec r}_n, s_n | T_H | d, s \rangle =
F_{m,N}(Q^2) e^{i {\vec q} \cdot {\vec r}_p }
\langle \vec {r}_{p}, s_{p}; \vec {r}_{n}, s_{n} \mid d, \vec {s} \rangle.
\end{equation}
Here  $m$ labels components in the $PLC$ ($N, N^*, N^{**}$).
The Green's function operator is diagonal in the hadronic mass
eigenstate basis :
\begin{equation}
\langle m, {\vec r}_p, s_p; n, {\vec r}_n, s_n | G |
m, {\vec r}_p\;', s_p; n, {\vec r}_n\;', s_n \rangle =
- {{ e^{i p_{m} | {\vec r}_p - {\vec r}_p' | }}\over{ 4 \pi
| {\vec r}_p - {\vec r}_p\;' | }}
\delta^{(3)}({\vec r}_n - {\vec r}_n\;').
\label{eq:heg}
\end{equation}
Here $p_{m}$ is the momentum of the $m'$th component of the proton wavepacket.
This is given by:
\begin{eqnarray}
p_{mz} & = & q - p'_{nz} \nonumber \\
p'_{nz} & =& {Q^2(1- {1\over x}) + M_{m}^{2} -M_{N}^2 \over 2q}.
\label{eq:energyconservation}
\end{eqnarray}
Here $p'_{nz}$ is the momentum of the neutron spectator in the intermediate
state (cf. discussion in sec. 5.4). The above expression is obtained using
energy-momentum conservation in intermediate states, valid in the
 semiclassical
approximation used here.  The eq.(\ref{D_3R}) shows that when  $|PLC\rangle$
state is modeled as a superposition with different masses, the longitudinal
momentum of nucleons in the intermediate state differs from the longitudinal
momentum of registered nucleons. This is another important difference of our
three resonance model\cite{FGMS93} from the model discussed in Ref.\cite{KOP}.

Here $\hat z$ is the direction of the $\gamma ^*$ momentum, $x$ is the Bjorken
variable, and we  take for simplicity $M_p=M_n=M_N$ and neglect the binding
energy of the deuteron. The quantity $M_m^2$ is the squared mass of the $m'th$
component of the wavepacket state.

The matrix element of the final-state interaction operator $T_S$ is
parameterized
according to eq.(\ref{UM}) with
\begin{equation}
\langle r_{p},  s_p; r_{n}, s_{n} |  T_S  |
m, {\vec r}_p, s_p; m', {\vec r}_n, s_n \rangle = U_{m,p}
\delta^{(3)}({\vec r}_p - {\vec r}_n).
\label{eq:heu}
\end{equation}
For the simplicity we neglect here the dependence of scattering  amplitudes on
momentum transfer (See however discussion in sec. 3.2.3).

We put all of the pieces together, insert complete sets of states, perform the
integrals over the delta functions, and arrive at the following expression for
the matrix elements:
\begin{eqnarray}
{\cal M}_0^s(\vec p_p,s_p;\vec p_n, s_n) &=& F_{p,p}(Q^2)
\int d^3 r_p d^3 r_n \,
e^{-i \vec p_p \cdot \vec r_p} e^{-i \vec p_n \cdot \vec r_n}
e^{i\vec q \cdot \vec r_p}
\langle r'_{p},s_{p};  \vec{4}_{n} s_{n} \mid d,\vec{s} \rangle,\\
{\cal M}_1^s(\vec p_p,s_p;\vec p_n, s_n) & = & i \, |\vec p_p|
\sum_m U_{p,m} F_{m,p}(Q^2) \int d^6 r \, e^{ -i
\left( {\vec k}_p + {\vec k}_n \right) \cdot {\vec r}_n}
\nonumber \\ && \quad \times
 e^{ i {\vec q} \cdot {\vec r}_p}
 {{ e^{i p_{p_m} | {\vec r}_p - {\vec r}_n | }}\over{ 4 \pi
| {\vec r}_p - {\vec r}_n | }}
\langle \vec{r}_{p}, s_{p}; \vec{r}_{n}s_{n} \mid d,\vec{s} \rangle,
\label{eq:hemcm}
\end{eqnarray}
We introduce the relative and center-of-mass coordinates according to
eq.({\ref{REL}), with this transformation, $d^3r_p d^3r_n = d^3R_{cm} d^3r$.
The integral over $d^3 R_{cm}$  leads to momentum  conservation:
$\vec q = \vec p_p + \vec p_n$ in the Lab. frame of the deuteron.
The remaining matrix elements are expressed purely in terms of the
relative coordinate, $\vec r$.

The non-relativistic deuteron wavefunction~\cite{Brown} can be written
in terms of the $s$- and $d$-state functions $u$ and $w$ as,
\begin{equation}
\langle \vec{r} \mid d, \vec{s} \rangle = \left[ \frac{u(r)}{r} + \frac{w(r)}{
\sqrt{8}r} S_{12}(\widehat r) \right] | \vec {s} \rangle,
\end{equation}
where the tensor operator
\begin{equation}
S_{12}(\widehat r)
 = {3 \vec \sigma_p \cdot \widehat r \vec \sigma_n \cdot \widehat r\over r^2}
- \vec \sigma_p \cdot \vec \sigma_n.
\end{equation}

As a result of the angular integration in $d^3r$ (in particular, the
integration over azimuthal angle ) $s_p + s_n = s$.  It is, thus,
convenient to define the quantities
\begin{equation}
{\cal M}_{0(1)}^{s}(\vec p_p, \vec p_n) \equiv
{\cal M}_{0(1)}^s( \vec p_p,s_p; \vec p_n, s_n).
\end{equation}
When $s=\pm1$ then $s_p=s_n=\pm \frac{1}{2}$.  When $s=0$, then
${\cal M}_{0(1)}^{s}$ is symmetric combination of $s_p=-s_n=\frac{1}{2}$
and $s_p=-s_n=-\frac{1}{2}$. The amplitudes for the  production of
final-state proton are
\begin{eqnarray}
{\cal M}_0^{s=\pm1}(\vec p_p, \vec p_n)
&=& 2\pi F_{pp}(Q^2) \left[ {\cal U}_0 + {\cal W}_0 \right],\\
{\cal M}_0^{s=0}(\vec p_p, \vec p_n)
&=& 2\pi F_{pp}(Q^2) \left[ {\cal U}_0 -2 {\cal W}_0 \right],\\
{\cal M}_1^{s=\pm1}(\vec p_p,\vec p_n)
&=& 2\pi \sum_m U_{p,m}F_{m,p}(Q^2) \left[ {\cal U}_m + {\cal W}_m \right],\\
{\cal M}_1^{s=0}(\vec p_p,\vec p_n)
&=& 2\pi \sum_m U_{p,m}F_{m,p}(Q^2) \left[ {\cal U}_m -2 {\cal W}_m \right],
\label{eq:deepn0}
\end{eqnarray}
where
\begin{eqnarray}
{\cal U}_0 &=& 2 \int_0^\infty dr \, r u(r) \frac{\sin p_n r}{p_nr},\\*
{\cal W}_0 &=& \sqrt{2} \int_0^\infty dr \, r w(r)
\left[ \frac{\sin p_n r}{p_n r} + 3 \frac{\cos p_n r}{p_n^2r^2} -
3\frac{\sin p_nr}{p_n^3r^3} \right],\\*
{\cal U}_m &=& -\frac{2}{4\pi} \int_0^\infty dr \, e^{ip_{p_m}r} u(r)
\frac{\sin qr}{qr},\\*
{\cal W}_m &=& -\frac{\sqrt{2}}{4\pi} \int_0^\infty dr \, e^{ip_{p_m}r}  w(r)
\left[ \frac{\sin qr}{qr} + 3 \frac{\cos qr}{q^2r^2} - 3\frac{\sin qr}{q^3r^3}
\right],
\end{eqnarray}
and where $q=|\vec q|$ and $p_n=|\vec p_n|$. Note that ${\cal W}_0 =0$ for
$\vec p_n = 0$.  At very high energies, $p_{p_m} \approx p_{p}$ is independent
of $m$, so that ${\cal U}_m$ and ${\cal W}_m$ become independent of $m$.
In this limit the matrix element ${\cal M}_1^s$ is proportional to
$\sum_m U_{p,m}F_{m,p} = 0$.  This is the limit of full color transparency.

The formulae of the conventional Glauber approximation can be obtained
from the above equations by retaining only the proton component in this sum.

\subsubsection{Modified Glauber Approximation Approach}

In this subsection we solve  the three-state model using the method developed
by Yennie \cite{YN}, K\"{o}lbig and Margolis \cite{KM} and Bauer \cite{BAU70}.
The major difference from the standard Glauber approximation is the
modification
of  the profile function by substituting $f_{NN}\rightarrow f_{nm}$,
 where $f_{nm}$
is the $m$-particle  production amplitude and the phase factor due
 to the difference
of the masses of particles $n$ and $m$\cite{BAU70}. This differs
from the previous
subsection and provides an opportunity to include a non-zero range
 of the interaction
(the slope $b_m$ of the amplitude $f_{nm}$).

Thus we can use equations of section 2, and modify the profile function
in eq.(\ref{GLS}) to calculate the $d(e,e'p)n$ processes within the Three
State Model:
\begin{equation}
\Gamma^N(b_p-b_n)  \rightarrow  {1\over 2}
\sum_{m=N,N^*,N^{**}}\int U_{p,m}{F_{m,p}\over F_{p,p}}
e^{{b_m\over 2}t}e^{i\vec k_t \cdot (\vec r_p-\vec r_n)}e^{-i\Delta_{m,p}
( z_p - z_n)} {d^2k_t\over (2\pi)^2},
\label{MPROF}
\end{equation}
where
\begin{equation}
\Delta_{m,p} \equiv |\vec p_m| - |\vec p_p| =
\Delta^0-\sqrt { M_N^2 + p_p^2 - M_m^2} - |\vec p_p| =
\Delta^0-{M_m^2-M_N^2\over 2|\vec q|}.
\label{ENCOR}
\end{equation}

In eq.(\ref{MPROF}) $b_m$ is the slope factor of the transition amplitude.
For the present  analysis we will assume that slopes of all transition
amplitudes are the  same as for $NN$ scattering.
The factor $e^{i\Delta_{m,p}(z_p - z_n)}$ accounts for the  phase shift
due to the different masses of PLC components. The additional factor
$F_{m,p}\over F_{p,p}$ accounts for the different coefficients for $|PLC>$
wave components. The main difference of eq.(\ref{MPROF}) from the formulae
of the Generalized Glauber approximation\cite{YN,KM,BAU70,BAR} is that
eq.(\ref{MPROF}) should satisfy the sum rules  eq.(\ref{SUM}) in  the point
where PLC is produced ($z_p-z_n=0$) and when $t\rightarrow 0$. Besides the
effects of the m-dependence of $t_{min}$  are effectively included in the
sum rule eq.(\ref{SUM}).

Using  modified profile functions in the formulae for the transition
operator  in eq.(\ref{OP})  we obtain:
\begin{eqnarray}
{\cal M} & = & F_{p,p}(Q^2)
\int d^3r \langle \vec{r}, s_{p}, s_{n} \mid d, \vec{s}\rangle
\left\{  e^{-i\vec p_i\vec r} \right. \nonumber \\
& & \left.  - {1\over 2i}\sum_m\int
{d^2k_t\over (2\pi)^2} U_{p,m} {F_{m,p}\over F_{p,p}}e^{{b_m\over 2}t}
\Theta(-z)\cdot e^{-i\Delta_{m,p}z}e^{-ip_i \cdot r}e^{ik_t \cdot b} \right\}.
\label{GGA2}
\end{eqnarray}
The  $z$-component for the momentum of knocked out proton in intermediate
state is :
\begin{equation}
p_{iz}^m =   p_{iz} + \Delta_{m,p},
\label{ZCOM}
\end{equation}
Finally we  obtain the following expressions for the deuteron decay function:
\begin{eqnarray}
D^{d}(q,p_p,p_n) = \rho^ {\vec s}_d(p_i,p_i) -
{1\over 2} \sum_{m=P,N^*,N^{**}} \int \rho^ {\vec s}_d(p_i,p^{m'}_i)\cdot
U_{p,m}{F_{m,p}\over F_{p,p}}\cdot e^{{b_m\over 2}t}
\cdot {d^2k_t\over (2\pi)^2} \nonumber \\
+ {1\over 16}\sum_{m,m'} \int \rho^ {\vec s}_d(p^m_{i1},p^{m'}_{i2})
U_{p,m}{F_{m,p}\over F_{p,p}}\cdot U^\dagger_{p,m'}
{F^\dagger_{m',p}\over F^\dagger_{p,p}}
\cdot e^{{b_m\over 2}t_1}e^{{b_m\over 2}t_2}
{d^2k_{t1}\over (2\pi)^2} {d^2k_{t2}\over (2\pi)^2}.
\label{D_3R}
\end{eqnarray}
where
$\vec p_i^{m'}~\equiv~\vec p_i^{m'}(p^m_{iz},p_{it}-k_t)$,
$\vec p_{i1}^m~\equiv~\vec p_{i1}^m(p^m_{iz},p_{it}-k_{t1})$,
$\vec p_{i2}^m~\equiv~\vec p_{i2}^m(p^m_{iz},p_{it}-k_{t2})$,
and density matrices $\rho^{\vec s}_d$ is same as defined in eq.(\ref{RHO_A}).

\section{Numerical Results}

To estimate  effects of final state interactions in the $d(e,e'p)n$ reaction
 we
compare the predictions of different models (considered in sections 2 and 3)
for the differential cross section and the tensor polarization asymmetry with
predictions of the plane wave Born impulse approximation.

The manifestations of color transparency depend strongly on the relative
importance of the screening (interference term between Born (Fig.1a) and FSI
amplitude (Fig.1b)) and double scattering terms (square of FSI amplitude).
To separate  different implications of CT it is necessary to distinguish
kinematical conditions where one of these  terms is dominant. Thus we analyze
the deuteron decay function in eq.(\ref{D_GA}) defined in Glauber
approximation,
for the  separate cases of unpolarized and polarized deuteron targets.

\subsection{(e,e'p) scattering on unpolarized deuteron target}

Consider first the  unpolarized deuteron case. We define the transparency $T$
as the ratio of the measured cross section (or calculated cross section
 with FSI)
to the one calculated in the PWIA \cite{Carol1,NE18,FLFS,GM,FSZ94}:
\begin{equation}
T(Q^2,p_p,p_n) \equiv {\sigma^{FSI}_{d(e,e'p)n}(Q^2,p_p,p_n)\over
\sigma^{PWIA}_{d(e.e'pn)}(Q^2,p_p,p_n)}.
\label{T1}
\end{equation}
Within the Glauber approximation, the cross sections are controlled by
the decay function in eq.(\ref{D_GA}) with  the unpolarized density matrix of
eq.(\ref{RHO_UP}).
In Fig.2 the  dependence of the transparency - $T(Q^2,p_p,p_n)$ on the
spectator neutron angle $\theta_n$ is presented for different values of the
spectator momenta - $p_n$ at fixed $Q^2~=~6~(GeV/c)^2$. The figure
demonstrates that the final state interaction is maximal for
$\theta_n\approx~90^0$ at smaller spectator momenta ($\leq~200~MeV/c$) and the
position of maximal FSI  shifts  to slightly  lower
spectator angles with increase of  $p_n$. The shift is the consequence of
kinematic factor $\Delta^0$ of  eq.(\ref{RF}) which accounts for the
fact that for rescattering of energetic particles the $E_n-p_n^z$ component is
conserved rather than $p_n^z$ (see eq.(\ref{lcom}) in section 5.4 below).
As follows from  eq.(\ref{lcom}) the maximal FSI effect occurs at
$x={Q^2\over 2 m q_0}=1$, which provides the zero value
of longitudinal momenta for intermediate spectator.
Below  we will refer to the exclusive kinematics corresponding to $x=1$ as the
perpendicular kinematics.

The fig.2 shows also that the final state interaction contributes differently
to the resulting cross section at different spectator momenta. At
$p_n~\leq~200~MeV/c$ (Fig2.a,b) FSI is dominated by the  screening effect
(the second term in eq.(\ref{D_GA})). At $p_n~\geq~300~MeV/c$ (Fig.2.c.d)
the double scattering term (the third term in eq.(\ref{D_GA})) plays the
dominant role at the $x\approx 1$  and it tends to diminish the screening
effects (dash-dotted curves in Fig.2). The sharp increase of the FSI with the
increase of spectator transverse momenta is the consequence of the decrease
of the values of  actual impact parameters contributing to  $NN$ rescattering.
Note that in calculations within the Glauber approximation the $pn$
scattering amplitude have been estimated using the relation:
\begin{equation}
f^{pn} = \sigma^{pn}_{tot}\cdot (i+\alpha_n)e^{b_n/2 t}
\label{para}
\end{equation}
where $\alpha_n=Re f/ Im f$ and parameters used  are those of
Refs.\cite{FSZ94,GM94,SILV,PRW}.

The above analysis suggests that  color transparency effects, which
 reduce the FSI
(rescattering amplitude), will have  different consequences for
different intervals
of spectator angles and momenta.

Results presented in Fig.3a indicate that transparency for the scattering
off the deuteron in perpendicular kinematics ($x=1$) should change quite
dramatically with $p_{n}$ from $T(p_{n}=0) \approx 0.97$ down to
$T(p_{n}=0.2~ GeV/c) \approx  0.5$. The decrease of $T$ shows  that FSI,
in this kinematical range, is mostly a screening effect (see Fig.2).
Note that current  NE-18 results\cite{NE18} for $T_d$ were obtained under
assumption that $T$ does not depend on $p_t$ in the studied momentum range
$p_t\le 300 MeV/c$. Therefore, to compare our results with these data it
is necessary to perform a new analysis of the data with a more  realistic
nuclear model where $T(p_t)$ decreases with $p_t$.

However at $p_{n}\geq 300~MeV/c$, $x=1$ the transparency $T$ starts to
 increase
with spectator momentum (Fig.3b), since in this kinematics (see Fig.2)
 the double
scattering term dominates.

As a consequence of different role of FSI in the  considered two kinematics,
CT effects will have  nontrivial manifestations. CT will increase $T$ as
compared to  GA predictions at $p_{n}\leq 200~MeV/c$, $x=1$
(see Fig.3a- curves "I,II,III") and will decrease $T$ as compared to GA
predictions at $p_{n}\geq 300~MeV/c$, $x=1$ (see Fig.3b- curves "IV,V").

The different pattern of CT effects for these two regions shows that
the ratio of the cross sections:
\begin{equation}
R(Q^2,p_{n1}, p_{n2}) = {\sigma(p_{n1}\approx 300,400 MeV/c)\over
\sigma(p_{n2}\approx 200 MeV/c)}
\label{R1}
\end{equation}
should be rather more sensitive to  CT phenomenon. This quantity is more
convenient to search for CT since it represents the ratio of directly
 measured
experimental quantities, and does not require   additional normalization to
the corresponding PWIA calculation (as in eq(\ref{T1})).  In Fig.4 we present
curves for the same kinematics as in Fig.3  where calculations
 accounting for
CT effects are normalized to the corresponding Glauber
 approximation calculations.
We calculate also the $Q^2$ dependence of the ratio-$R$ (curves labelled by
the boxes). One can see that CT can modify R by as much as $30\%$ for $Q^2$ as
low as $6-10~(GeV/c)^2$. Note that this occurs in the kinematical region where
competing  nuclear effects in the deuteron are small and
are under good theoretical
control (see sec. 5).

In Fig.5 we compare the predictions of different CT approximations for
transparency defined as in eq.(\ref{T1}). Comparison of calculations within
the three resonance model shows that the Green function formalism
 (see sec.3.2.2)
predicts smaller DWIA cross section(Fig.5 - solid lines labeled by boxes) than
the modified GA approach (see sec.3.2.3). This difference is caused
by the slope
of the $NN$ scattering amplitude in the modified GA approach.  Including this
effect allows the transverse internal nucleon momenta in the deuteron to be
smaller in the modified GA approach. This is a qualitatively new feature.
 However,
the effects of the finite range of interaction do not exceed $10\%$ in the
 discussed
kinematical range $p_n\leq~300MeV/c$. In the case of CT, for spectator momenta
($p_n\leq~200MeV/c$), neglecting the slope of the soft rescattering amplitude
predicts approximately the same effects as modified GA  (Fig.5a - dashed
curves
labeled by boxes). For larger spectator momenta ($p_n\geq~300MeV/c$) neglect
 of
slope for rescattering amplitude enhances CT effect (Fig.5b - dashed curves
 labeled
by boxes).  This is because for higher spectator momenta  rescattering  occurs
effectively at smaller distances than in the case of the modified  GA. These
considerations demonstrate that up to deuteron internal nucleon momenta
$\sim 200MeV/c$  one can neglect the finite sizes of the rescattered nucleons,
while at higher momenta when internuclear distances become comparable with
nucleon
size one needs to account for the form factors of the  interacting nucleons.

The energy dependence of final state interactions
of the $d(e,e'np)$   reaction involving a
spinless deuteron target has been recently discussed in
Ref.~\cite{ANIS}.  That paper studies ratios of cross section integrated over
$p^n_t$ for various values of Bjorken x
 and uses the triple pomeron  mechanism of diffraction, which is not applicable
for the energies considered in this paper.

\subsection{(e,e'p) scattering off the polarized deuteron}

The possibility to employ a polarized deuteron target to investigate color
coherent effects is quite tempting. Use of different  polarization states
enhances the role of the $d$-wave component of the deuteron wave function
and so that smaller space-time intervals are probed. This results in tagging
of the PLC at the early stage of its evolution  to a normal hadron.

For numerical estimates we consider the tensor polarization $T_{20}$
measurable
in electrodisintegration of the polarized deuteron. We define $T_{20}$ as:
\begin{equation}
T_{20} \equiv {1\over 3}\left(\sigma(1,1) + \sigma(1,-1) -
2\cdot\sigma(1,0)\right)
= \sigma_{ep}\cdot D^d_{20}(q,p_p,p_n)\cdot\delta(q_o - M_d - E_p - E_n),
\label{T_20}
\end{equation}
where  $\sigma(s,s_z)~\equiv~{d\sigma^{\vec s,s_z}\over dE_{e'} d\Omega_{e'}
d^3p_p}$, $s$ and $s_z$ are the spin and it's  $z$ component of the deuteron.
The decay function is defined according to Eq.(\ref{D_GA}) with the tensor
polarization density function as:
\begin{eqnarray}
\rho^d_{20}(k_1,k_2) & \equiv &  {1\over 3}\left(\rho_d^{-{i\over\sqrt{2}}
(a_x+ia_y)}(k_1,k_2) + \rho_d^{{i\over \sqrt{2}}(a_x-ia_y) }(k_1,k_2) -
2\cdot \rho_d^{i\cdot a_z}(k_1,k_2)\right)\nonumber \\
& =  &
\left[3{k^2_{2z}\over k_2^2} - 1\right]{u(k_1)w(k_2) \over \sqrt{2}} +
\left[3{k^2_{1z}\over k_1^2} - 1\right]{u(k_2)w(k_1) \over \sqrt{2}}
\nonumber \\
\nonumber \\
& & +\left( {3\over 2}
\left[{(k_1\cdot k_2)\cdot ((k_1\cdot k_2) - 3k_{1z}k_{2z})\over k_1^2k_2^2}
 + {k^2_{1z}\over k_1^2} +{k^2_{2z}\over k_2^2}\right] - 1\right)
w(k_1)w(k_2).
\nonumber \\
\label{RHO_20}
\end{eqnarray}
Eqs.(\ref{RHO_A}) and (\ref{VK}) are used in obtaining eq.(\ref{RHO_20}).

For further calculations we define an asymmetry as the ratio of cross
sections from  a tensor polarized and   unpolarized deuteron:
\begin{equation}
A_d(Q^2,p_p,p_n) \equiv
T_{20} / {d\sigma^{unp}\over dE_{e'}d\Omega_{e'} d^3p_p} =
{D^d_{20}(q,p_p,p_n)\over D^{unp}_{d}(q,p_p,p_n)},
\label{asm}
\end{equation}
where the unpolarized decay function - $D^{unp}_{d}(q,p_p,p_n)$ is defined
according to eq.(\ref{D_GA}) and (\ref{RHO_UP}).

Before considering  manifestations  of CT for  $T_{20}$ and $A_d$ it is
worthwhile  to examine some properties of the tensor polarization density
matrix eq.(\ref{RHO_20}). Considering the scattering from unpolarized
deuterons
(section 4.1) demonstrates that FSI dominates for nearly perpendicular
 kinematics:
$\vec p_s \perp \vec q$. Thus in  Fig.6 we compare $\rho^d_{20}(k,k)$
at $k_z=0$
with the  corresponding unpolarized deuteron density function
$\rho_d(k,k) = u(k)^2+w(k)^2$ calculated with different (Paris and Bonn) $NN$
potentials. Fig.6 illuminates several remarkable properties of the tensor
polarization density functions. First, it practically coincides with
 unpolarized
density function for $k\approx~300~MeV/c$, where
 $u(k)\approx~-{w(k)\over \sqrt{2}}$.
This property reflects the fact that the $s$- partial
 wave in the deuteron falls with
$k$, and changes sign at $k\approx~400~MeV/c$, while
 $d$-wave grows with $k$ from
negative minimum at $k\approx~100~Mev/c$. Thus in
 some range of momenta, the
influence of the $d$-state is  comparable (or larger)
 than that of the $s$-state
(for a more detailed discussion see \cite{Brown}).
 Note that we follow here the
convention of Refs.\cite{Paris,Bonn} which define
 the deuteron wave function so
that $w(k)<0$ at small $k$. The opposite convention, corresponding to w(k)$>$0
is adopted e.g. in Refs. \cite{Brown,FS88}). This equality of polarized
and unpolarized density functions means that in this kinematical range the
asymmetry calculated according to eq.(\ref{asm}) will be close to unity,
for perpendicular kinematics provided there is no final state interaction
(i.e. in the  plane wave Born approximation). This prediction is  practically
independent of the $NN$ potential used to compute the deuteron wave function
(see Fig.6). Besides, we will see in sec.5 that other effects such as the EMC
effect, off-shell and relativistic effects are small for our kinematics. So
deviations of the asymmetry from unity  originate predominantly from the
influence of final state interactions.

The next important feature of the tensor polarization density function is that
it decreases with decrease of internal momenta (Fig.6), and we find that
average
momenta in the integrals for the final state interactions in the perpendicular
kinematics are $100-150~MeV/c$. Therefore, within the framework of the
conventional eikonal approximation large effects of FSI are expected, since
rescattering occurs at rather small internucleon distances $\sim  1.5~fm$.
 This
dominance of FSI by small internucleon distances indicates that the effect
of PLC
evolution  to a normal size nucleon will be reduced. Therefore one should
expect
larger sensitivity to CT effects.

In Fig.7 we present a three-dimensional plot  of the asymmetry $A_d$ as a
function
of the spectator momenta and the  polar angle, calculated within the PWIA
(Fig.7a)
and the Glauber approximation (Fig.7b). This figure clearly demonstrates
 the large
influence of final state interactions
within the framework of Glauber approximation for
perpendicular kinematics with $p_n~\approx~250-350~MeV/c$ and
$\theta_n~\approx~60-90^0$.

In Fig.8 the $Q^2$ dependence of $A_d$  is calculated using the PWIA, Glauber
approximation and CT models considered in text, for  different values of
spectator transverse momenta. The sensitivity to CT effects is noticeable.

In Fig.9 the $Q^2$ dependence for both the ratio $R(Q^2,p_{n1},p_{n2})$
defined according to eq.(\ref{R1}) and  asymmetry $A_d$ defined according to
eq.(\ref{asm}) calculated for different values of $\Delta M^2$, which
 determines
the coherence length  within the QDM prediction of color transparency. Here we
restrict $Q^2$ to the  range available at CEBAF in the near future
($Q^2 \sim 6~(Gev/c)^2$)\cite{CHEN}.
One can see that already in the $E_e=4-6 GeV$ CEBAF run, it is possible
to obtain important constraints on the parameters of CT.

\section{Theoretical uncertainties}

In this section we consider  several theoretical issues which might influence
the reliability of our interpretation of the measured cross sections.

\subsection {Relativistic motion of target nucleons}

The fact that  high energy processes  develop along the light-cone can be
 taken
into account within the framework of the light-cone mechanics\cite{FS81,FS88}.
The light-cone calculation of processes involving a deuteron target
is more straightforward than that using nonrelativistic quantum mechanics. The
parameter which characterizes the importance of relativistic motion of
nucleons
in the deuteron is the difference between the deuteron internal momentum
defined in the laboratory reference frame ($p_n$) and in the  light-cone
reference frame \cite{FS81}:
\begin{equation}
k_n = \sqrt{ {m^2+p^2_{nt}\over \alpha(2-\alpha)} - m^2},
\label{KLC}
\end{equation}
where $\alpha= {p_{n-}\over p_{d-}}$ is the light-cone fraction of deuteron
momentum carried by spectator nucleon.
In the perpendicular kinematics of present interest $\alpha\approx 1$ and it
follows from eq.(\ref{KLC}) that $k_n\approx p_{nt}$. Therefore  the expected
relativistic effects of nucleon motion in the deuteron are insignificant
corrections ($\sim{\cal O}(1-\alpha)^2)$. Another potentially possible source
of difference between nonrelativistic and light-cone descriptions is the
nucleon spin rotation effect\cite{FS83}. However it was demonstrated in
Ref.\cite{FS88} that in the case of perpendicular kinematics this difference
is negligible once again. Note also, that relativistic effects are even less
important for the FSI, since integrals in the FSI amplitude
(see e.g. eq.(\ref{D_GA})) are sensitive to rather smaller values of the
deuteron internal momenta.

\subsection{Uncertainties in knowledge of the deuteron wave function}

Possible uncertainties in the calculation of the $d(e,e'p)n$ processes with
polarized and unpolarized targets are small and under control for the
kinematics considered here because nucleon momenta in the deuteron do not
exceed $300-350MeV/c$ and FSI terms are sensitive to smaller deuteron
internal momenta.  Another reason why our processes are less sensitive to
the uncertainty of the deuteron wave function, is that we consider
observables which are the  ratios of experimental quantities in (nearly)
similar kinematical conditions (see eqs.(\ref{R1}) and (\ref{asm})).

\subsection{Off shell effects}

The target nucleons are bound in the deuteron; the square of their
four momentum is not the square of their mass. The nucleons are
off-shell. One needs to estimate the influence of this effect to
calculate the cross section of  eq.(\ref{CRS}). We estimated  the
uncertainties due to these effects by considering  the Born amplitudes
of models which account differently for off-shell effects in $\sigma_{ep}$
\cite{DEFOR,FS88}. The differences are very small, less than one  percent or
so.

One of the uncertainties originates from the deformation of the bound nucleon
wave function. This deformation has been calculated in Ref.\cite{FS85} for
processes dominated by PLC. A similar  effect arises within Skyrmion models
of the two nucleon interaction\cite{AIS}. The major effect is the suppression
of the probability for PLC in bound  nucleon due to the color screening
phenomenon\cite{FS88}. The influence of this effect can be estimated by
rescaling the deuteron wave function for a nucleon with momentum $k$ by the
factor:
\begin{equation}
\delta(k) =  \left(1 + \Theta(Q^2-Q^2_0)\cdot (1-{Q^2_0\over Q^2})\cdot
{{k^2\over m} + 2\epsilon_d\over \Delta E}\right)^{-1},
\label{delta}
\end{equation}
where $\epsilon_d$ is the deuteron binding energy and $\Delta E$($\approx
 0.6~GeV$)
is the parameter which characterize  bound nucleon excitations in the
deuteron. The
$Q^2$ dependence accounts for the presumed dominance of PLC in bound
 nucleons for
sufficiently large momentum transfer $Q^2\geq Q^2_0\approx
2(GeV/c)^2$\cite{FSZ94}.

In Fig. 10 we present  calculations of the ratio defined in
 eq.(\ref{R1}) (Fig.10a)
and the asymmetry defined by eq.(\ref{asm}) (Fig.10b) which
include the influence
of the PLC suppression effect of Eq.(\ref{delta}). Figures demonstrate that
considered uncertainties are on the level of $5\%$ and $10\%$ in the case of
unpolarized and polarized measurements respectively.

\subsection{Semiclassical \ approximation \ for \ FSI \ in \ the
 \ momentum \ \
space - Feynman diagram approach}

When calculating FSI for the  $d(e,e'p)n$ reaction we substituted $\Theta(-z)$
contribution in eq.(\ref{GA_M3}) by ${1\over 2}$ (cf. Eq.(\ref{GA_M4})).
This replacement allows to represent the FSI amplitude (Fig.1b) as the
convolution integral of the deuteron wave function and the on-energy shell
$NN$ scattering amplitude in the momentum space. The aim of this subsection is
to substantiate our approximations. We start by analyzing the Feynman diagram,
Fig.1b, which describes the FSI. Our interest is in the kinematics where the
momentum of the nucleon-spectator - $p_n$ is small:
 ${p^2_{n}\over m_n^2}\ll 1$.
In this case it is reasonable to consider the nonrelativistic motion of the
nucleon in the deuteron: ${\vec p'^2_{n}\over m_n^2}\ll 1$, where
$p'_{n}$ is the momentum of spectator in the intermediate state
($\vec {p'}_n =
\vec p_n - \vec k$, $k$ is transferred momentum in the amplitude of FSI - see
Fig1.b). So it is legitimate to evaluate the loop integral by taking a residue
over the energy  of the spectator nucleon in the intermediate state
- ${p'}_{n0}$.
(For nonrelativistic motion of the nucleon this is the only pole in the lower
part of the complex plane in the variable ${p'}_{n0}$.) Neglecting
systematically all the terms $\sim {\vec p'^2_{n}\over m_n^2}$ as compared to
$1$ we obtain:
\begin{equation}
{{\cal M}^{FSI}\over <p'|J^{em}_\mu|p>}  =
 -{(2\pi)^{{3\over 2}}\over 2}
\int \psi(p_n'){f^{NN}(p'_n-p_n,s)\over p'_{nz}-p^0_{nz}+i\epsilon}
{d^3p'_n\over (2\pi)^3}
\label{inter}
\end{equation}
Here the momentum $\vec q$ is chosen to be in the $z$ axis direction,
$s=(p^\mu_d+q^\mu)^2$ and:
\begin{equation}
p^0_{nz}\equiv (x-1)m{q_0\over \sqrt{Q^2+q_0^2}} = p_{nz}-\Delta^0 =
p_{nz} - (E_n-m){M_d+q_0\over \sqrt{Q^2+q_0^2}}
\label{lcom}
\end{equation}
where $p_{nz}$ is the $z$-component  of measured spectator momentum.
The last term of eq.(\ref{lcom}) accounts for real kinematics of the considered
process. The amplitude $f^{NN}$ in the eq.(\ref{inter}) is normalized
according
to $Im f^{NN}(s,t=0) = \sigma_{tot}$ and the deuteron wave function normalized
as $\int\psi^2(p_n)d^3p_n=1$.

Eq.(\ref{lcom}) reflects an
important property of two-body high energy processes
in which
the variable $k_-=k_0-k_z$ (fig.1b) but not $k_z$ is small. This  dynamics
is easily accounted for in light-cone mechanics of deuteron \cite{FS81}.

It is easy to demonstrate that eq.(\ref{inter}) is equivalent to the usual
Glauber approximation if off-energy shell effects are neglected in $f^{NN}$.
Really, if we use the Fourier transform of the deuteron wave function as:
\begin{equation}
\psi(p'_n) = {1\over (2\pi)^{3\over 2}}\int \psi(r) e^{-i\vec p'_n \vec r}d^3r
\label{furier}
\end{equation}
the integral in eq.(\ref{inter}) over $p'_{nz}$ differs from $0$ for $z<0$
only.
This is the factor $\Theta(-z)$ in the coordinate space representation.

We shall neglect in the following analysis the dependence of $f^{NN}$ on
$p'_{nz},p_{nz}$, since they are small for two body processes at high energies
and will ignore off shell effects for small momenta of spectators.
Then the integral
over $p'_{nz}$ can be evaluated by deforming the contour integral over
 $p'_{nz}$
into the lower part of the complex plane. To calculate this integral we
need to know
analytic properties of the wave function of deuteron. For this we use the
 conventional
parametrization of deuteron wave function calculated with  Paris\cite{Paris}
 and
Bonn\cite{Bonn} $NN$ potentials:
\begin{equation}
\psi(p) =  \sum_j {C_j\over p^2+m_j^2}
\label{fm}
\end{equation}
where $\sum_j C_j~=~0$ and $s$ and $d$-waves differ by coefficients  $C_i$.
Substituting  $\psi(p_n')$ in eq.(\ref{inter}) from eq.(\ref{fm}) we obtain:
\begin{eqnarray}
{{\cal M}^{FSI}\over <p|J^em_\mu|p>} & = & -{(2\pi)^{{3\over 2}}\over 2}\sum_j
\int {d^2p'_{n\perp}\over (2\pi)^2}\int  {d^2p'_{nz}\over (2\pi)} \times
\nonumber \\
& & {C_j\over (p'_{nz}+i\sqrt{p'^2_{n\perp}+m_j^2})
(p'_{nz}-i\sqrt{p'^2_{n\perp}+m_j^2})}
{f^{NN}\over p'_{nz}-p^0_{nz}+i\epsilon}
\nonumber \\
 & = & {i\over 2}\sum_j \int {d^2p'_{n\perp}\over (2\pi)^2}f^{NN}
\left[{C_j\over \tilde p_n^2+m_j^2} -
{iC_j(p^0_{nz}-i\sqrt{p'^2_{n\perp}+m_j^2}\over
2\sqrt{p'^2_{n\perp}+m_j^2}(\tilde p_n^2+m_j^2)}\right]
\label{pole1}
\end{eqnarray}
where $\tilde p_n\equiv (p^0_{nz},p'_{n\perp})$. Separating the real and
imaginary parts inside of $[...]$ and using eq.(\ref{fm}) one obtains a
factor ${1\over 2}$ which we used in eq.(\ref{GA_M3}) and additional term
neglected in the above calculations:
\begin{eqnarray}
{{\cal M}^{FSI}\over <p|J^em_\mu|p>} & = & {i\over 2}
\int {d^2p'_{n\perp}\over (2\pi)^2}f^{NN} \left[{\psi(\tilde p_n)\over 2}
- {ip^0_{nz}\over 2}\sum_j {C_j\over \sqrt{p'^2_{n\perp}+m_j^2}
(\tilde p_n^2+m_j^2)}\right] \nonumber \\
& = & {i\over 4} \int \psi(\tilde p_n) f^{NN} {d^2p'_{n\perp}\over (2\pi)^2}
\times\left\{1 - i\beta\right\},
\label{pole2}
\end{eqnarray}
where
\begin{equation}
\beta = {p^0_{nz}\int {d^2p'_{n\perp}\over (2\pi)^2}f^{NN}
\sum {C_j\over \sqrt{p'^2_{n\perp}+m_j^2}\cdot(\tilde p_n^2+m_j^2)}\over
\int \psi(\tilde p_n) f^{NN} {d^2p'_{n\perp}\over (2\pi)^2}}.
\label{pole3}
\end{equation}
The size of $\beta$ is a measure of the accuracy of replacing the $\Theta(-z)$
function by the factor ${1\over 2}$. Eqs.(\ref{pole2}), (\ref{lcom}) and
(\ref{pole3}) show that within the Glauber approximation the contribution
from the factor $\beta$ decreases with spectator longitudinal momentum
$p_{nz}$
and is practically negligible for perpendicular kinematics. However in the
case of the modified Glauber approximation for the
three resonance model
 (sec.3.2.3)
the longitudinal momentum of intermediate state does not  coincide with
the longitudinal momentum of external detected particle.
Due to this difference, the  value of $\beta$ in the three resonance model
is determined at $p'_{nz} = p^0_{nz}-\Delta_{m,p}$, where $\Delta_{m,p}$
is defined according to eq.(\ref{ENCOR}). Therefore the contribution of
$\beta$ will be larger than within the conventional Glauber approximation.
Note that contribution of $\Delta_{m,p}$  in soft FSI  are effectively
included
into the sum rule eq.(\ref{SUM}).

It follows from eq.(\ref{pole2}) that the contribution of  $\beta$ factor to
the interference of Born and FSI amplitude is further suppressed because the
real part of  $f^{NN}$ amplitude is small ($\sim 0.2$ of the imaginary part)
and because the Born term corresponding fig1.a is real. The contribution of
$\beta$ factor to the double scattering term is $\sim\beta^2$
(cf. eq.(\ref{pole2})).

In Fig.11 we demonstrate the $Q^2$ dependence of the ratio defined by
eq.(\ref{R1}) (Fig.11a) and the asymmetry defined by eq.(\ref{asm}) (Fig.11b)
within the conventional Glauber approximation and the modified Glauber
approximation for the three resonance model. It follows from this calculation
that including $\beta$ within the three resonance model of modified GA brings
the predictions of this model and QDM approximation (where we included the
$\Theta(-z)$ factor explicitly in coordinate space) closer together.

\subsection{Meson exchange currents (MEC) and $\Delta$-isobar contributions}

Estimates of contributions from meson exchange currents corresponding to
diagrams similar to Fig.1c are rather controversial at large $Q^2$ since these
terms are very sensitive to the assumed t-dependence  of the meson-nucleon
 vertex
form factors. These form factors are not obtained  from theory; instead they
 are
used as fitting parameters\cite{Aren}. On the other hand  the restriction to
$x={Q^2\over 2mq_0}\approx 1$ and $Q^2\geq~1~(GeV/c)^2$ strongly suppresses
 the
sea quark content in nucleons. Including mesonic components is one way to
 treat
the  anti-quark content, so that a suppression of the sea can be considered
 as a
hint of the suppression of MEC at large $Q^2$ and $x\ge 1$. This argument
 relies
on the Bloom-Gilman duality \cite{BG} between the structure function of a
 nucleon
at $x\rightarrow 1$ and contribution of resonances and on the fact that this
hypothesis describes reasonably near threshold data for deep inelastic $eN$
scattering. Besides, the diagram corresponding to Fig.1c  contains an extra
${1\over Q^2}$ factor as compared to the  Born (Fig.1a) and the  FSI (Fig.1b)
diagrams.

The $\Delta$-isobar contribution is expected to be small also in the
kinematical range suitable for studying CT effects, since the
$\gamma^*N\rightarrow\Delta$ transition form factor decreases more rapidly
with $Q^2$ than the $N^*$ transition form factors\cite{STL}. Another reason
 for
the suppression of the contribution of $\Delta$'s is that the
$\Delta N\rightarrow NN$ amplitude is predominantly real and decreases rapidly
with energy (since it is dominated by pion exchange) whereas the FSI effects
 we
study are determined by imaginary part of the soft rescattering amplitude.

In  fig.12  we compare predictions of the Glauber approximations
of this paper with results of calculations of the model of Ref.\cite{AREN0}.
\footnote{We are thankful to W.~Leidemann for making available his calculations
within the approach of Ref.\cite{AREN0} for the kinematics  discussed in this
paper.}
Calculations has been performed in the kinematics $Q^2=1~(GeV/c)^2$,
$q_0\sim 400-500~MeV/c$ and spectator momentum $p_n=400~MeV/c$. Within model
of Ref.\cite{AREN0} the contribution of meson currents and isobars in the
kinematics where $NN$ rescattering dominates is small ($\sim 6\%$ for MEC
and $\sim 4\%$ for isobar contribution).  Comparision of our approach and that
of Ref.\cite{AREN0} hints that Glauber approximation is applicable for
$Q^2$ as small as $1~(GeV/c)^2$. Note that calculation according to
Ref.\cite{AREN0} can be considered as the upper limit for the contribution of
meson currents. Theoretical analyses of current  data (cf. Ref.\cite{BFS}) and
calculations of Ref.\cite{BB} indicate that the contributions of meson
exchange  currents have been strongly overestimated  previously.

\subsection {Charge exchange contribution and spin dependence of the elementary
 amplitude}

In this analysis we have neglected contributions of terms in which the
large longitudinal momentum is first transfered to the neutron which then
converts to a proton via a charge exchange process. Since the charge exchange
reaction $np \rightarrow pn$ is dominated by the pion exchange its amplitude
is predominantly real. Therefore it mostly contributes to the double
interaction term, not to the screening term. We can estimate its relative
contribution as
\begin{equation}
{\sigma_{charge exchange} \over \sigma_{double}} \approx
{\sigma_{en,elastic} \over \sigma_{ep,elastic}}
{d \sigma^{np \rightarrow pn}/dt \over
d \sigma^{pn \rightarrow pn}/dt}_{\mid t \sim 0.05 (GeV/c)^2}
\end{equation}
Using the data from Ref. \cite{chargeexch}, we estimate that this correction
is of the order of $9\%$ for $Q^2 \sim 2 GeV^2$ and it decreases rapidly with
$Q^2$, approximately as $Q^{-4}$. For the case of leading neutron production,
the relative contribution of the charge exchange is larger by a factor
$\sigma^2_{ep}/\sigma^2_{en} \leq 5$.  Note that we expect transparency
effects to cut down the charge exchange  just as they reduce elastic
scattering. This is the chiral transparency effect discussed in
Ref.~\cite{FMS92}; so this charge exchange effect does not affect our
conclusions  concerning the sensitivity of the $d(e,e'p)n$ reaction to
color coherent effects.

Note also that we neglected another effect of dependence of the elastic $pn$
amplitude on the parallel or anti-parallel nature of the helicities of the
colliding nucleons. If we consider, for certainty, magnetic transitions,
the rescattered nucleons will predominantly be in the   antiparallel
(parallel) state for $\lambda_d=1(0)$. Using the current data on the
$\sigma^{tot}_L(pn)$ difference \cite{sigmal}, we estimate that this
effect leads to a correction to the rescattering amplitude which is
$\leq 3\%$ for $Q^2 \sim 2 (GeV/c)^2$ and  $\leq 1\%$ for $Q^2 \sim
4 (GeV/c)^2$. This effect is somewhat more important for $A_d$
calculations. We will consider the effects discussed here in more detail
elsewhere.

\subsection{Factorization approximation}

In the DWIA form of cross section (eq.(\ref{CRS})) we assume that the
electron-nucleon cross section is not changed by the distortion of
kinematics due to final state interaction. This commonly used assumption
for the distorted wave impulse approximation  allows us to factorize the
cross section in the form of eq.(\ref{CRS}). However computing the
unfactorized form of cross section  for the  deuteron is quite straightforward
and the authors will discuss it elsewhere. For small nucleon momenta within
deuteron this effect is small anyway.

\section{Summary}

We have demonstrated that selecting perpendicular kinematics for the
spectator:
$p_{n}\sim 200-400~MeV/c$, $x\sim 1$  significantly increases the effect of
struck nucleon rescattering off the spectator nucleon. Thus varying $p_{n}$
allows the selection of smaller than average internucleon distances in the FSI.
If $p_n$ is small the final state interactions cause a screening effect, and
the effect of including the color transparency is to increase the
computed cross section over the value obtained using the Glauber approximation.
But for larger values of $p_n$, the conventional final state interactions
enhance the cross section. In this case, the influence of color transparency
is to suppress the computed cross sections. Based on these observations,
we suggest measuring  the ratio of $d(e,e'p)n$ cross sections, where
in one case FSI is dominated by screening effect and in the second case by the
double scattering. This ratio  is more sensitive to the effects of color
transparency. Our calculations of this ratio, using several CT models, predict
$20-40\%$  CT effects for  $Q^2\sim 4-10~(GeV/c)^2$.

High sensitivity of the $d(e,e'p)n$ cross section to the final state
interaction of sufficiently energetic struck nucleon  is one the of
preconditions for looking for color transparency phenomena. This
condition is valid in  the  case of high energy electrodisintegration
of a polarized deuteron with production of a spectator under the
perpendicular kinematics discussed above. We suggest measuring the
tensor asymmetry in $\vec d(e,e'p)n$ reaction, where conventional
Glauber approximation predicts a huge influence of FSI. The dominance
of deuteron $d$-state wave in the polarized density matrix restricts the
essential internucleon distances to  $\sim~1.5~fm$, and suppresses the
PLC evolution to a normal size nucleon. The CT models we consider give
predictions of $50-100\%$ effects at $Q^2\sim 4-10~(GeV/c)^2$.

To estimate the possibility for the  unambiguous investigation of CT
effects with deuteron target we  considered a number of theoretical issues
which might affect  the reliability of the theoretical calculation of the
$d(e,e'p)n$ reaction. The analysis shows that under our perpendicular
kinematic conditions one can confine the theoretical uncertainties to the
level of $\sim 5\%$ and $\sim 10\%$ for unpolarized and polarized measurements
respectively.

Comparison with models explicitly accounting for meson currents and isobars
shows that in the  perpendicular kinematics preferable for  searching for
CT effects  the deuteron can be considered as a clean detector with a low rate
of noise.

\section{Acknowledgments}

This work was supported in part by the U.S.A. - Israel Binational Science
Foundation Grant No. 9200126 and by the U.S. Department of Energy under
Contract Nos. DE-FG02-93ER40771 and DE-FG06-88ER40427.

\newpage
\centerline{Figure Captions}

\begin{itemize}

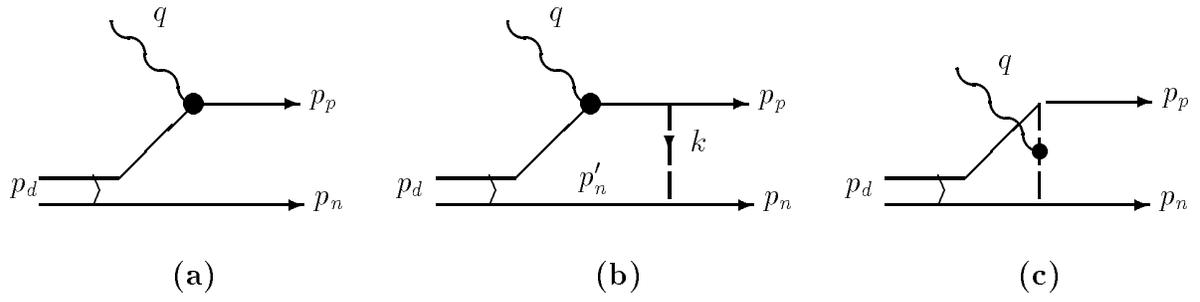
\item [ Figure 1.]  Graphs for d(e,e'p)n scattering. (a) - Born (PWIA)
approximation; (b) - final state interaction contribution,
 where the broken line accounts for all $NN$ multiple scatterings;
(c) - meson exchange current contribution.

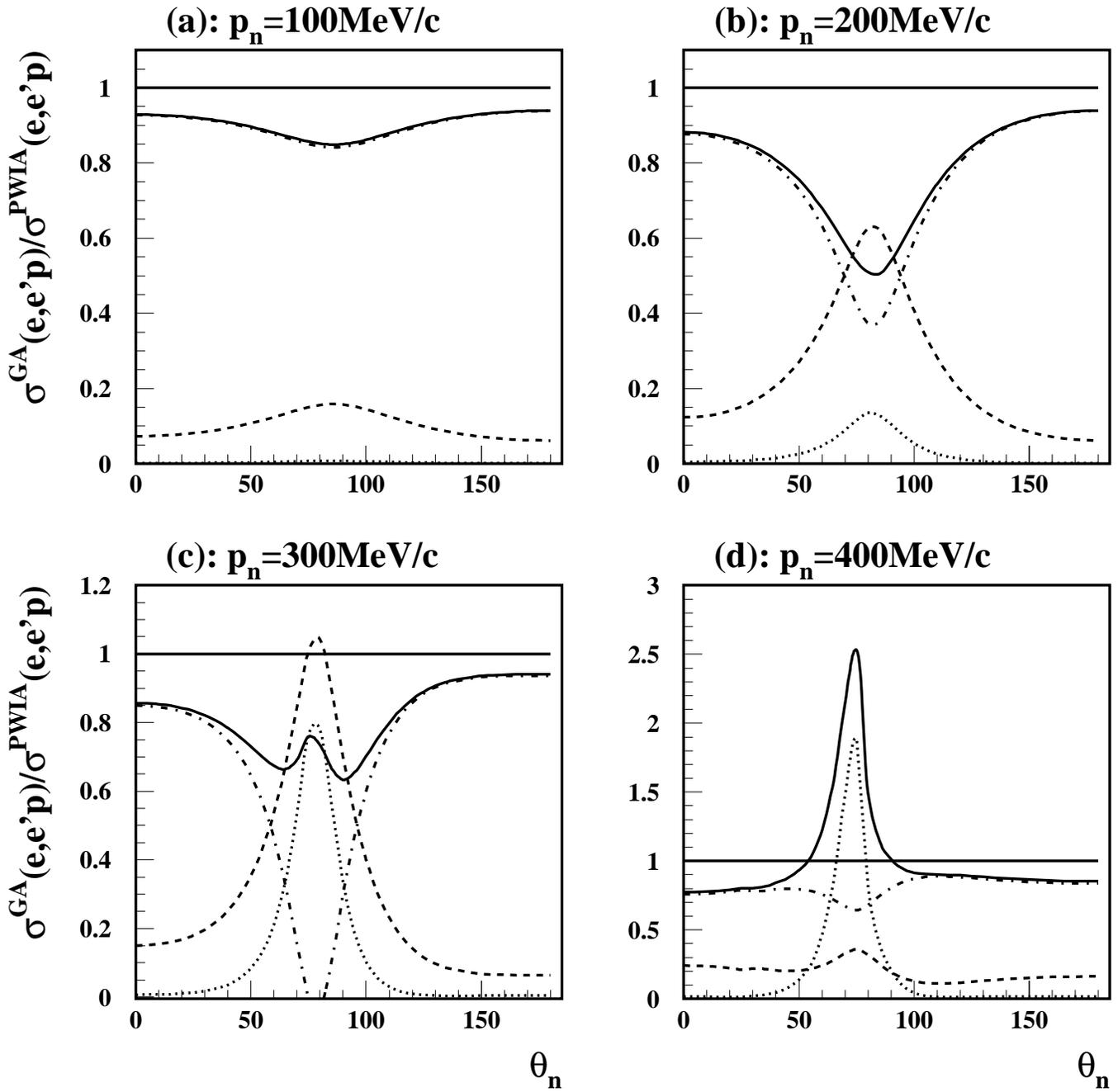
\item [ Figure 2.] The dependence of  transparency - $T$ (eq.(\ref{T1}))
on the spectator neutron angles, for different values of neutron momenta:
(a) $p_n~=~100~MeV/c$, (b) $p_n~=~200~MeV/c$, (c) $p_n~=~300~MeV/c$ and
(d) $p_n~=~400~MeV/c$. The solid-straight line is the Born approximation,
dashed line - contribution of interference between Born and FSI amplitude,
dash-dotted - effects of screening - is the sum of Born and interference
terms, dotted line - contribution of the double scattering term, solid-line
- is the sum of all terms in the cross section. $Q^2=6~(GeV/c)^2$.

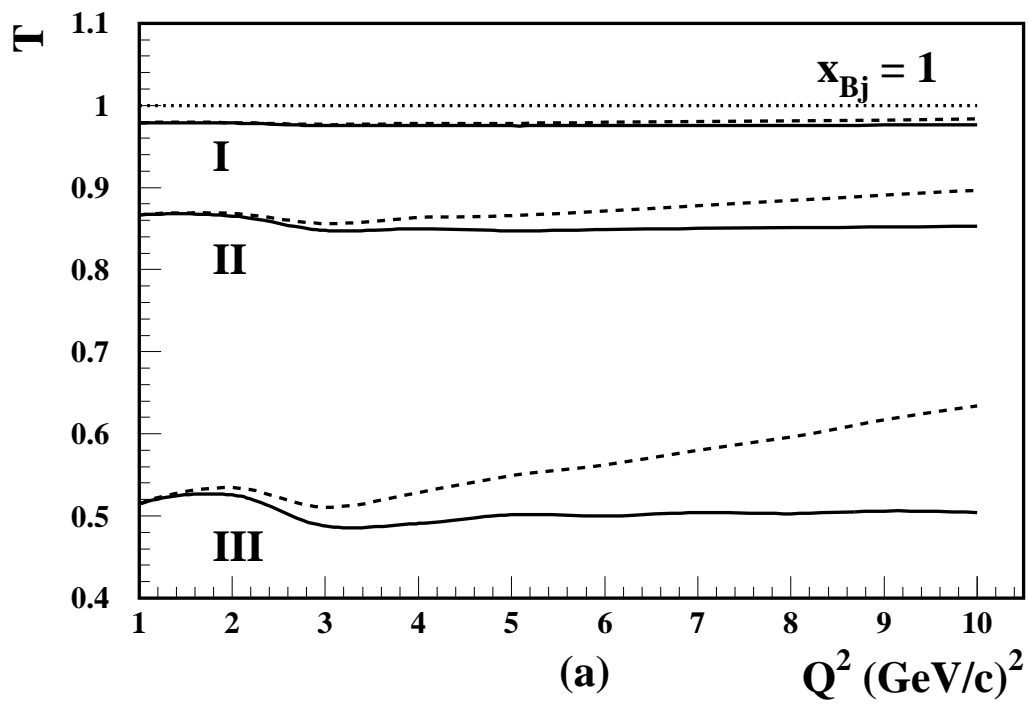
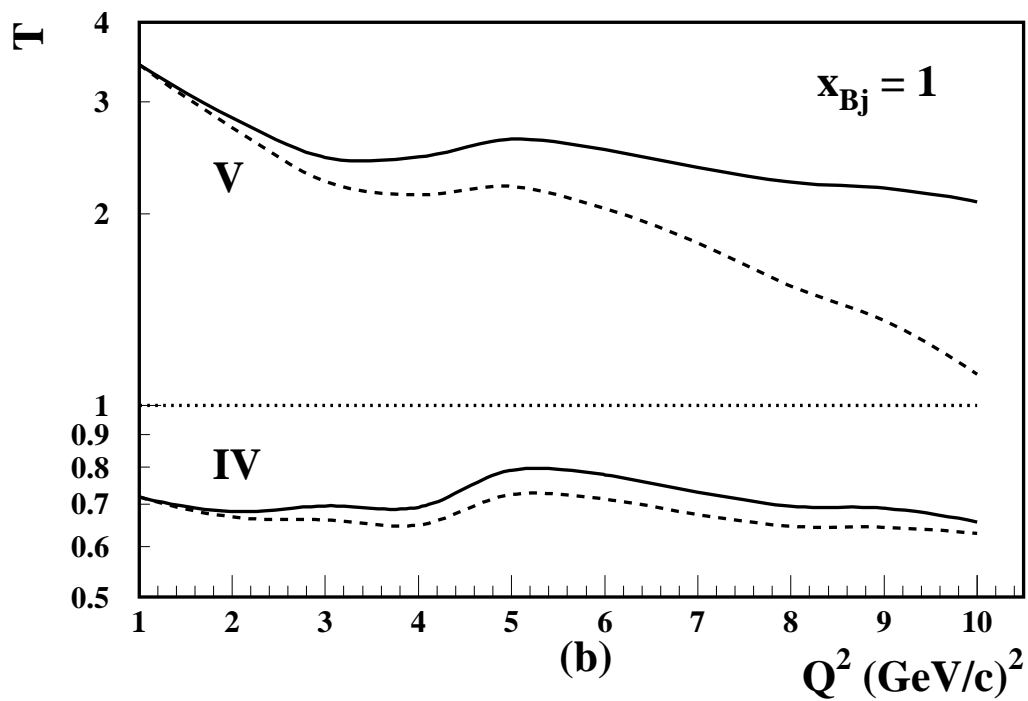
\item [ Figure 3.] $Q^2$ dependence of transparency $T$ defined according
to eq.(\ref{T1}) at $x=1$. Solid line - GA prediction, dashed line - CT
prediction within QDM approximation, with $\Delta M^2=0.7~GeV^2$ and dotted
line PWIA prediction. Curves labelled "I" are the calculations at $p_n=0$,
"II" - at $p_n=100~MeV/c$, "III" - at  $p_n=200~MeV/c$ "IV" - at
$p_n=300~MeV/c$, and "V" - at  $p_n=400~MeV/c$.

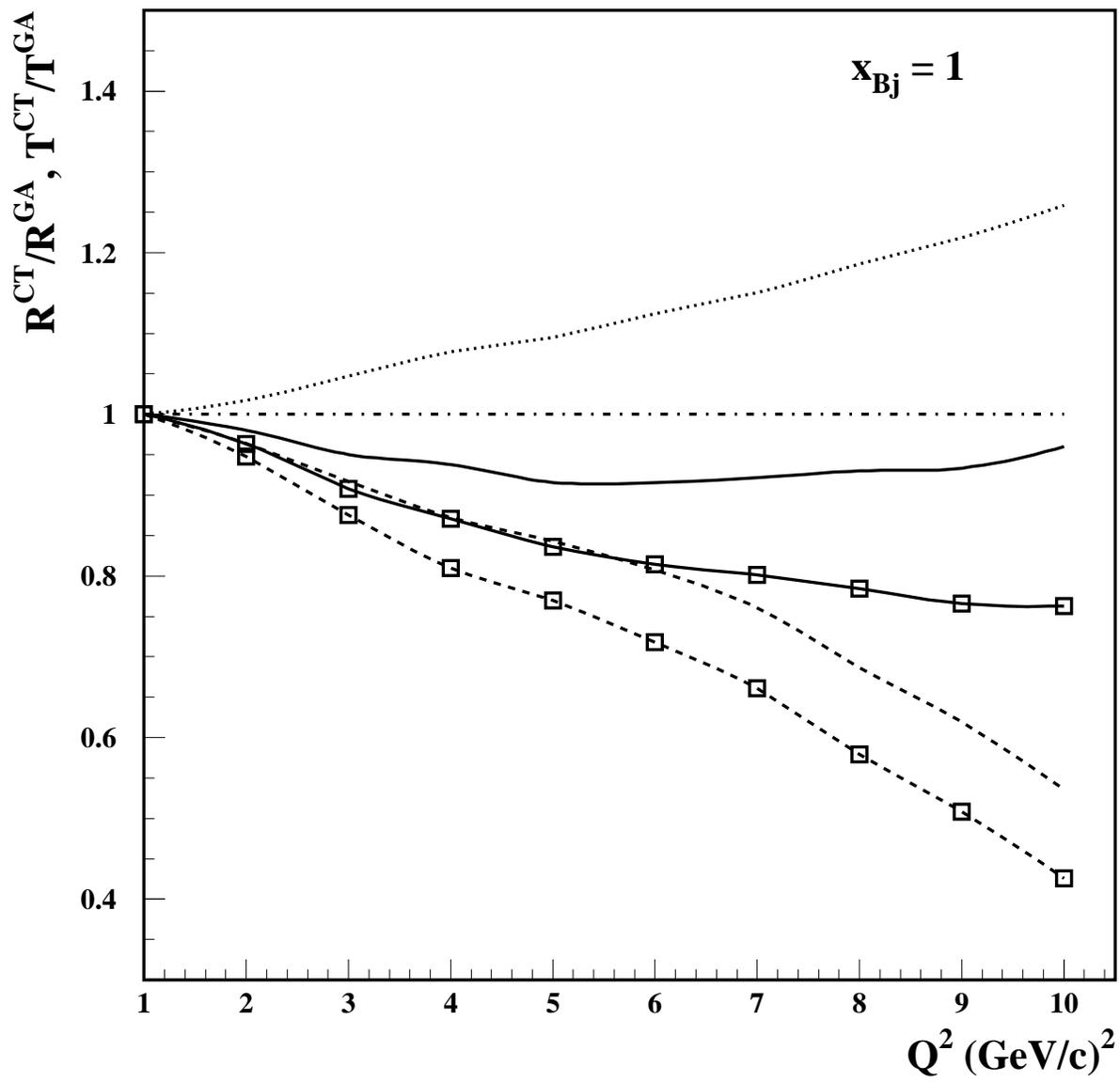
\item [ Figure 4.] $Q^2$ dependence of transparency $T$ defined according
to eq.(\ref{T1}) normalized on GA calculations: dash-dotted line GA
approximation, dotted line - QDM prediction at $p_n=200~MeV/c$, solid line
- QDM prediction at $p_n=300~MeV/c$ and dashed line - QDM prediction at
$p_n=400~MeV/c$. The curves labelled by boxes correspond to $Q^2$ dependence
of QDM predictions for $R(Q^2,p_{n1},p_{n2})$ defined according to
 eq.(\ref{R1}),
normalized on GA calculations. Solid line with "boxes" - $p_{n1} = 300~MeV/c$,
$p_{n2} = 200~MeV/c$ and dashed line with "boxes" - $p_{n1} = 400~MeV/c$,
$p_{n2} = 200~MeV/c$. QDM predictions calculated with $\Delta M^2=0.7~GeV^2$.
$x~=~1$.

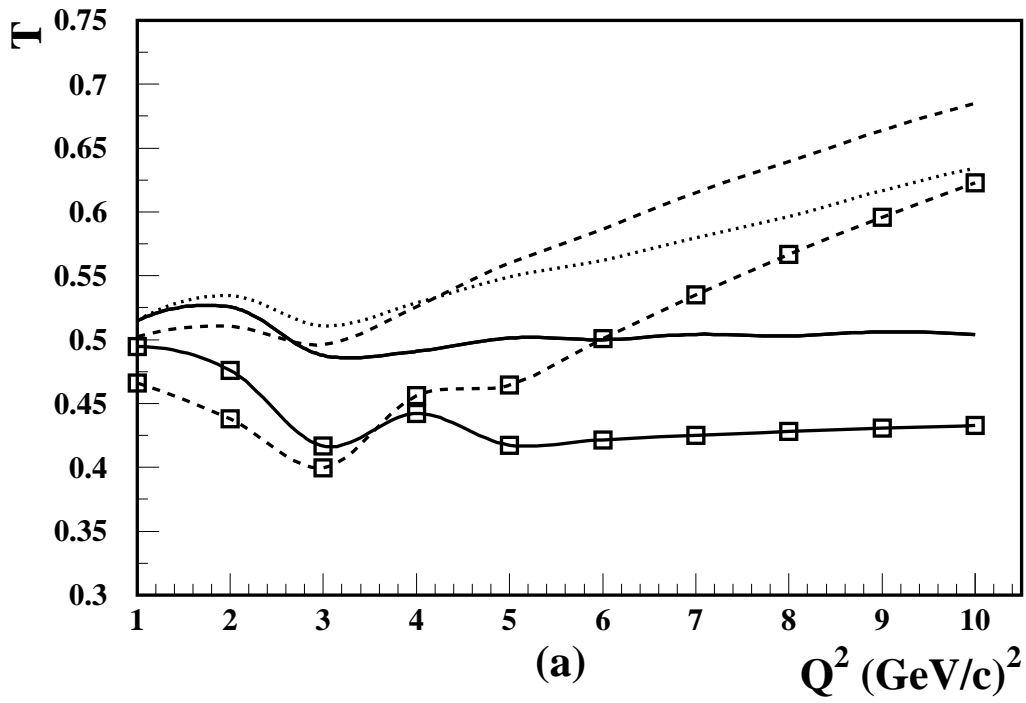
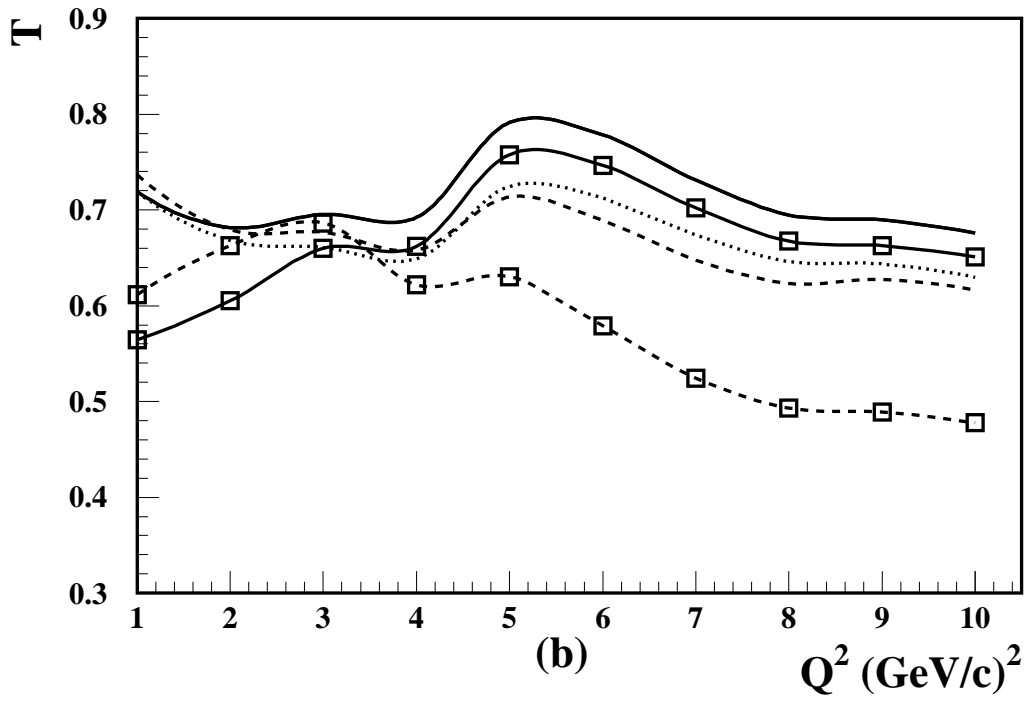
\item [ Figure 5.] $Q^2$ dependence of transparency $T$  at $p_n=200~MeV/c$ -
(a) and $p_n=300~MeV/c$ - (b), for $x~=~1$. The solid-line - prediction
within the GA, dotted line - QDM approximation with $\Delta M^2=0.7~GeV^2$,
dashed line three resonance model within modified GA, with parameters
$M_{N^*}=1.4~GeV$, $M_{N^{**}}=1.8~GeV$, $\epsilon=0.17$, ${F_{N,N}\over
F_{N,N^{**}}} = 1.0$ and ${F_{N^*,N}\over F_{N,N^{**}}} = 3.1$. Solid and
dashed
lines labelled by boxes are calculations within the GA and three resonance
 model
of CT within Green function method and zero-range $NN$ scattering
approximation.
The parameters for three resonance model are same as above.

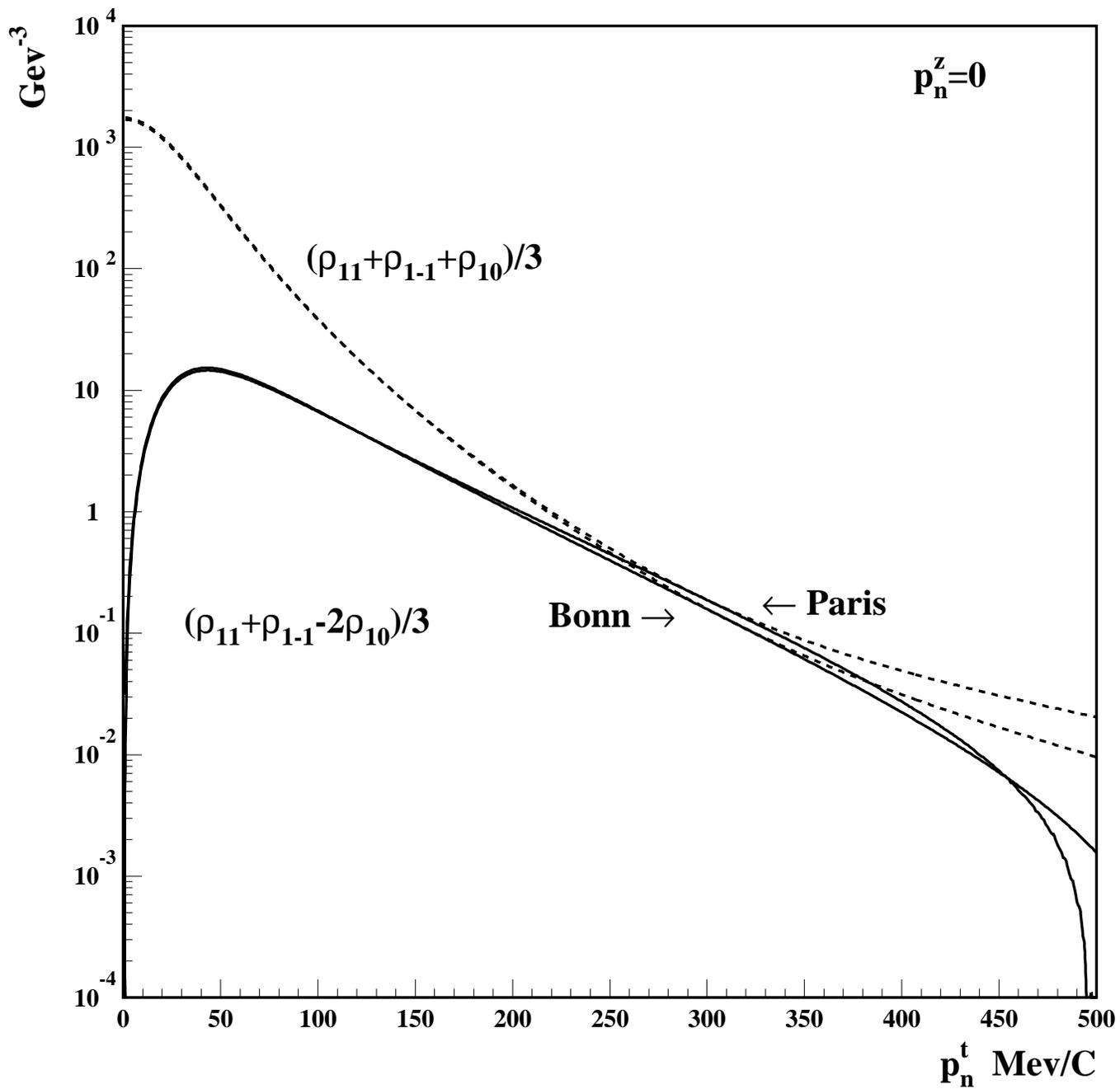
\item [ Figure 6.] Transverse momentum dependence of unpolarized
 density matrices
- $\rho_d(k,k) = u(k)^2+w(k)^2$  (eq.(\ref{RHO_UP})) - dashed line and
 polarized
density matrices - $\rho_{20}^d(k,k)$ (eq.(\ref{RHO_20})) - solid line,
 at $p_n^z=0$.

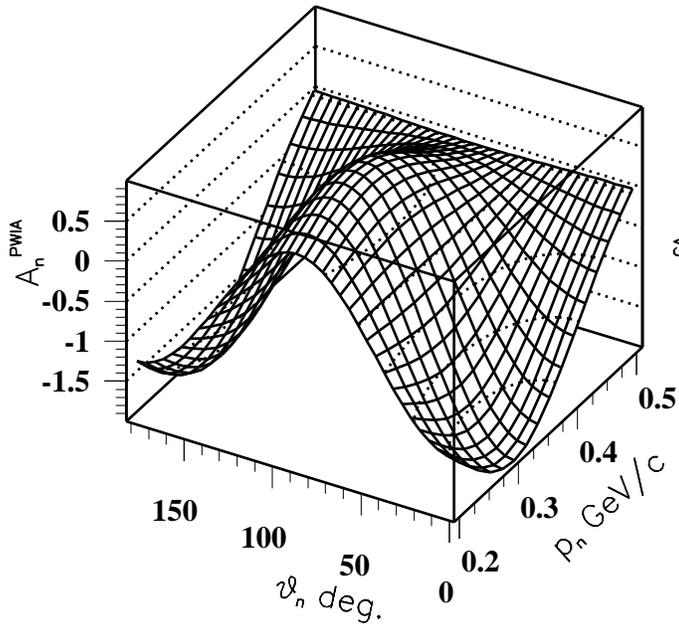
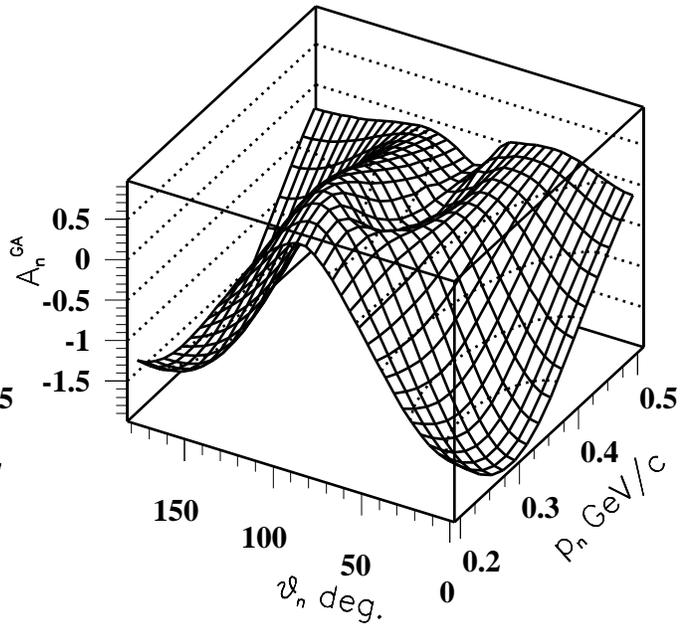
\item [ Figure 7.]  The $(\theta_n,p_n)$ dependence od asymmetry defined  by
 Eq.(\ref{asm}). (a) - Born (PWIA) approximation, (b) - Glauber approximation.
$Q^2=6~(GeV/c)^2$.

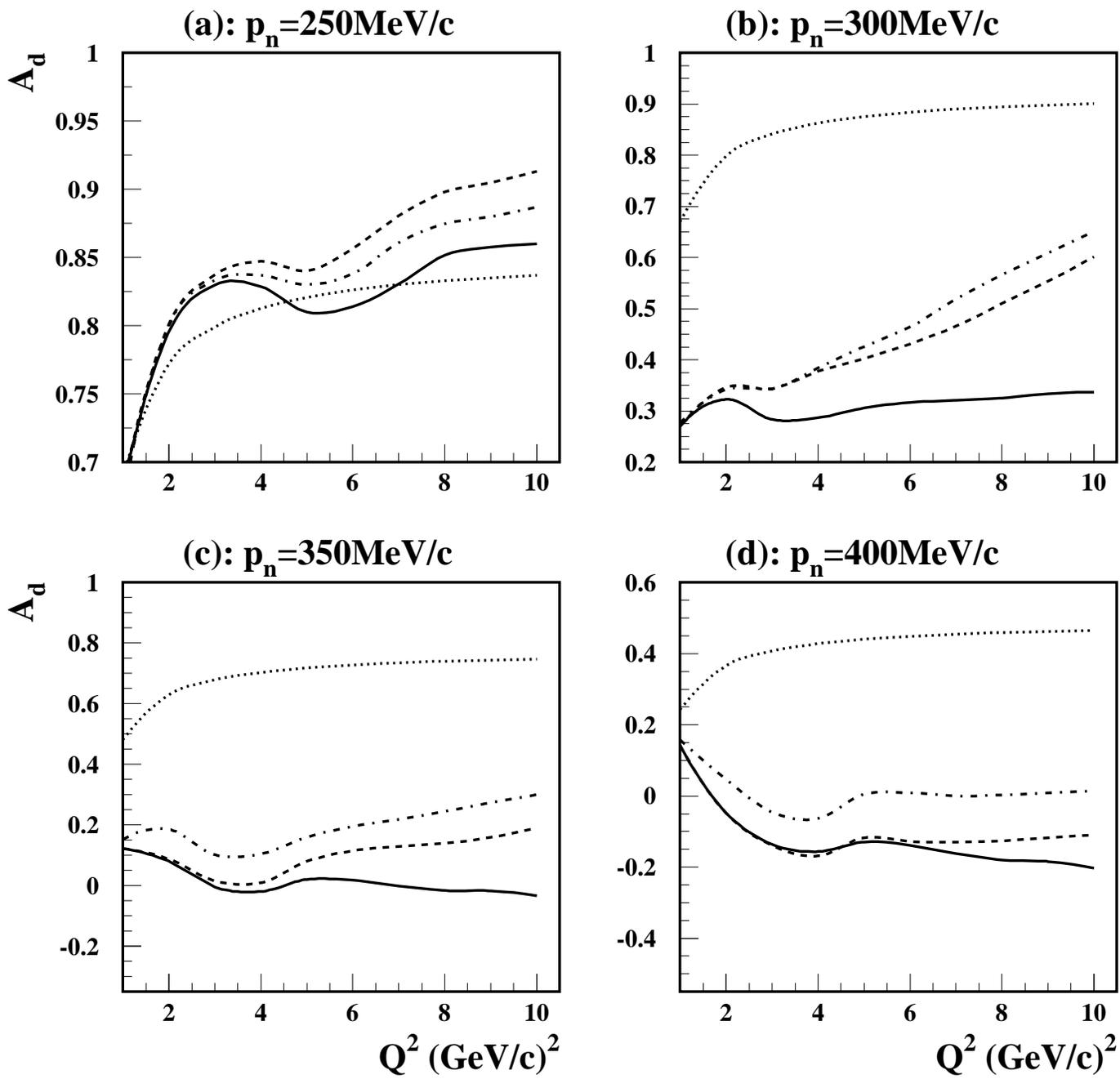
\item [ Figure 8.] $Q^2$ dependence of asymmetry defined according to
eq.(\ref{asm}), at $x~=~1$. Solid line - Glauber approximation, dashed line -
QDM of color transparency, dash-dotted line - the three resonance model of
 color
transparency within the modified GA. Straight-dotted line - Born
approximation.
(a) - $p_n=250~MeV/c$, (b) - $p_n=300~MeV/c$,
(c) - $p_n=350~MeV/c$  and (d) - $p_n=400~MeV/c$.
QDM and three resonance model parameters are same as in Fig.5.

\item [ FIgure 9.]  $Q^2$ dependence the ratio $R(Q^2,p_{n1},p_{n2})$
defined according to eq.(\ref{R1}), at $p_{n1}=300~MeV/c$ and
$p_{n2}=200~MeV/c$ - (a) and asymmetry $A_d$ defined according to
eq.(\ref{asm}) at $p_n=300~MeV/c$ - (b). $x~=~1$.
Solid line - Glauber approximation, dashed line - QDM approximation
with $\Delta M^2=0.7~GeV^2$, dotted line - $\Delta M^2=0.9~GeV^2$ and
dash-dotted line - $\Delta M^2=1.1~GeV^2$

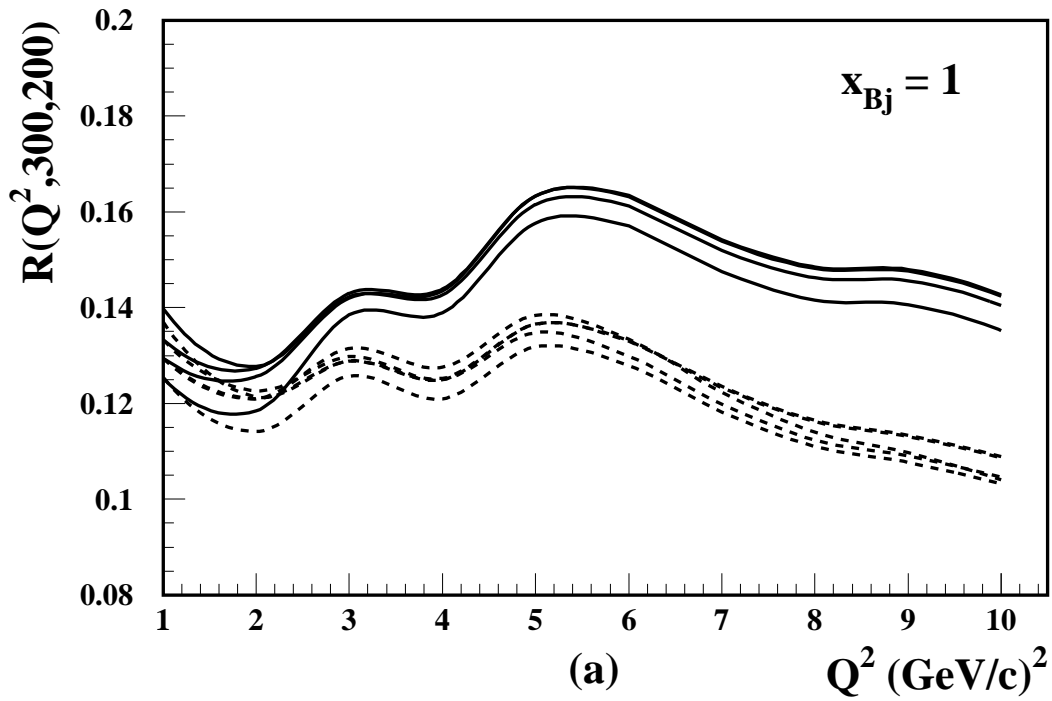
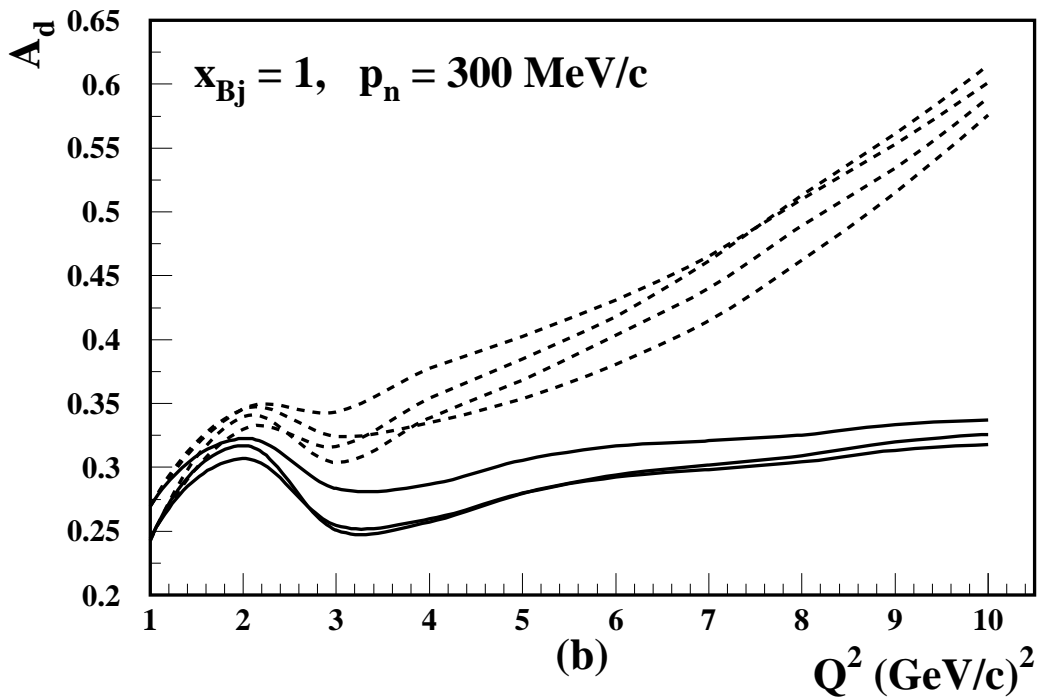
\item [ Figure 10.] The theoretical uncertainties in determining of $Q^2$
dependence of transparency $R(Q^2,p_{n1}=300,p_{n2}=200)$ - (a) and asymmetry
$A_d(p_n=300)$, at $x~=~1$. Solid-line - glauber approximations,
dashed - lines QDM approximations. Upper, down  solid, dashed
lines at $Q^2=1~(GeV/c)^2$ - are the GA and QDM predictions with Paris
and Bonn wave functions respectively. The curves inside corresponds
to predictions with Paris light cone wave function, color screening effects
(which according to eq.(\ref{delta}) appears at $Q^2>2~(GeV/c)^2$) and
predictions with different approximation for $\sigma_{en}$
(eq.(\ref{CRS}). QDM calculated with $\Delta M^2=0.7~GeV^2$.

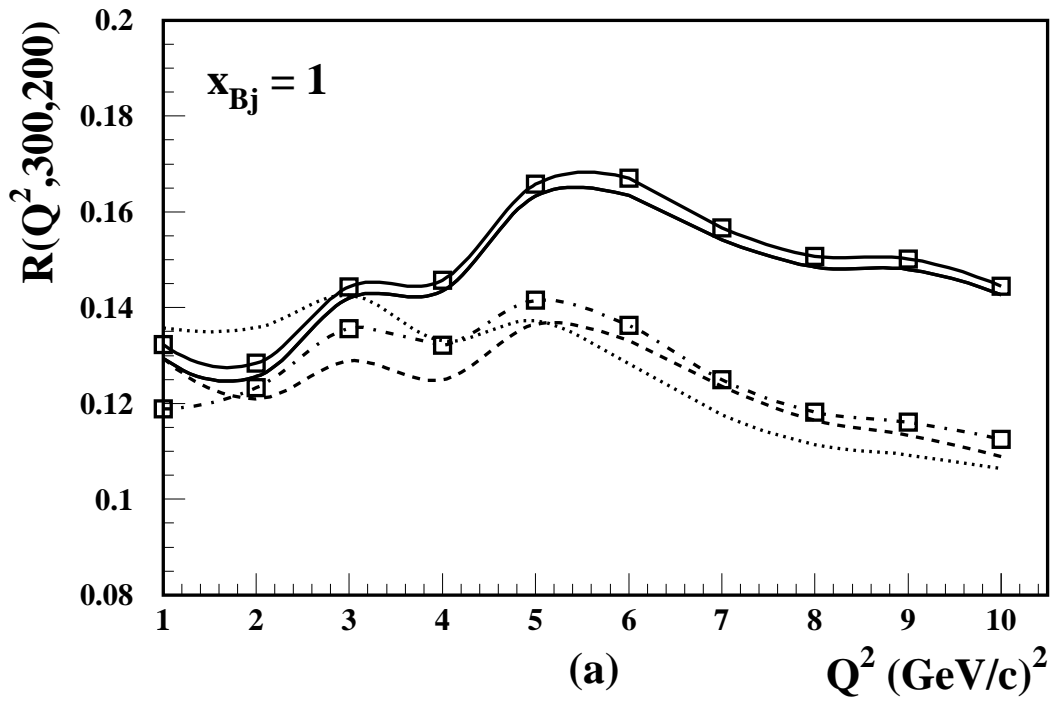
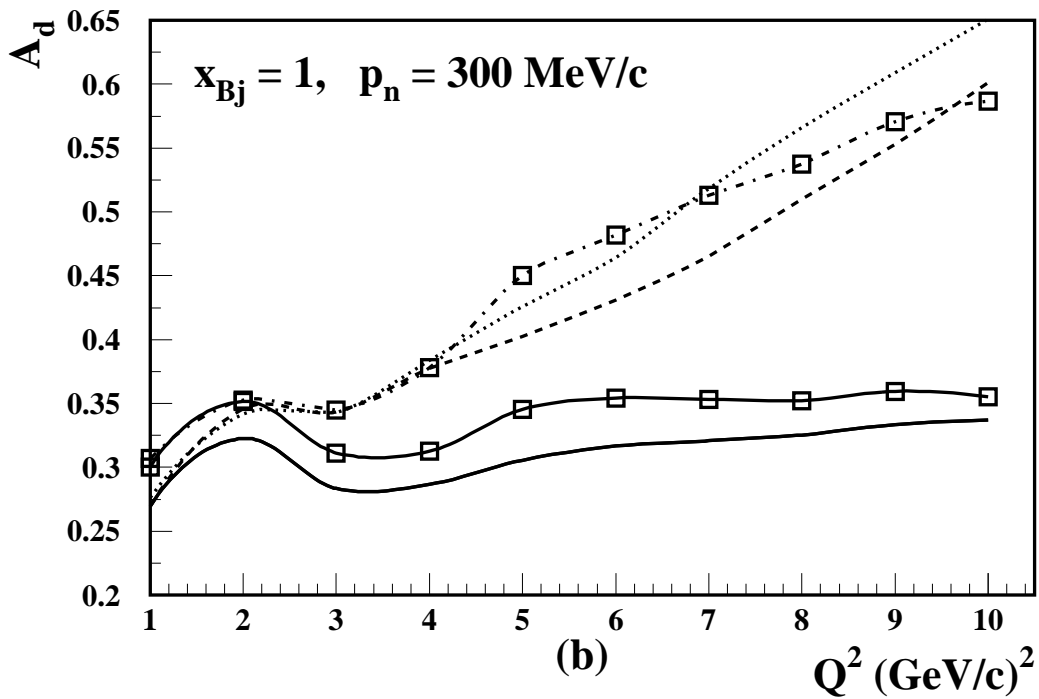
\item [ Figure 11.]  $Q^2$ dependence the ratio $R(Q^2,p_{n1},p_{n2})$
defined according to eq.(\ref{R1}), at $p_{n1}=300~MeV/c$ and
$p_{n2}=200~MeV/c$ - (a) and asymmetry $A_d$ defined according to
eq.(\ref{asm}) at $p_n=300~MeV/c$ - (b). $x~=~1$. Solid line -
Glauber approximation, dashed line - three resonance model of CT, within
modified Glauber approximation, doted line - QDM approximation of CT, solid
line with "boxes" - Glauber approximation with $\beta$ factor considered in
eq.(\ref{pole2}) and dashed line with "boxes" - three resonance model of CT,
within modified Glauber approximation and with  $\beta$ factor corresponding
to higher mass excitation in the intermediate state.
QDM and three resonance model parameters are same as in Fig.5.

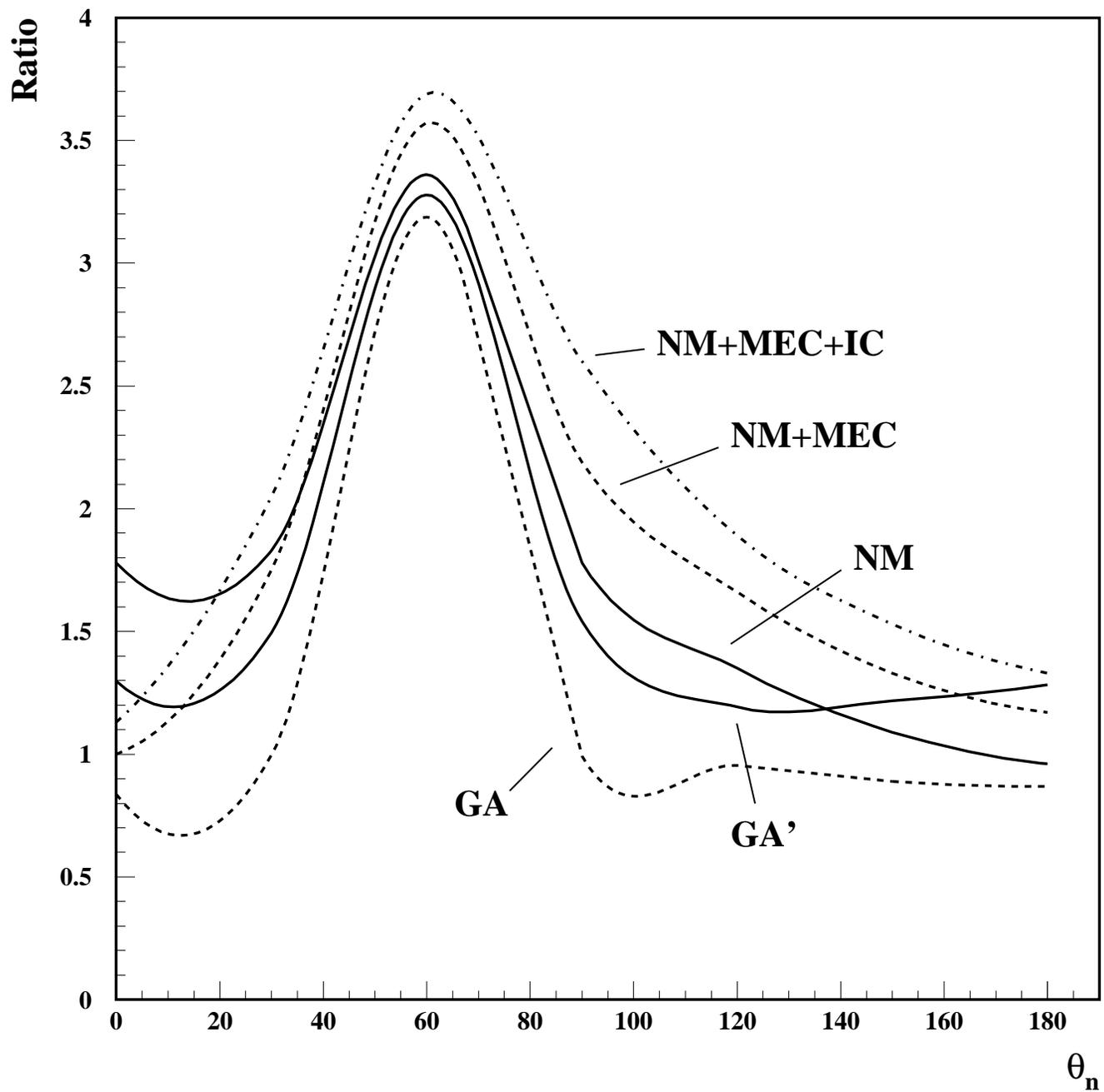
\item [ Figure 12.] The $\theta_n$ dependence of ratio   $R$.
Curve marked by "GA"   shows the  ratio of $d(e,e'p)n$ cross section
calculated within Glauber approximation, without factor of $\beta$
(eq.(\ref{pole3}) to the PWIA cross section, "GA$^\prime$" - same as above,
with Glauber cross section calculated with factor $\beta$,  "NM" - ratio of
cross section with multiple scattering  and Siegert operator to the Born cross
section, which also accounts for the direct productions of slow neutrons,
"NM+MEC" - "NM" calculation with meson exchange currents,  and "NM+MEC+IC"
-"NM" calculation with  meson exchange currents and isobar contribution.
The last three curves are those of the  calculation of  Ref.\cite{AREN0}
for   $p_n~=~400~MeV/c$ and $Q^2~=~1~(GeV/c)^2$.
\end{itemize}
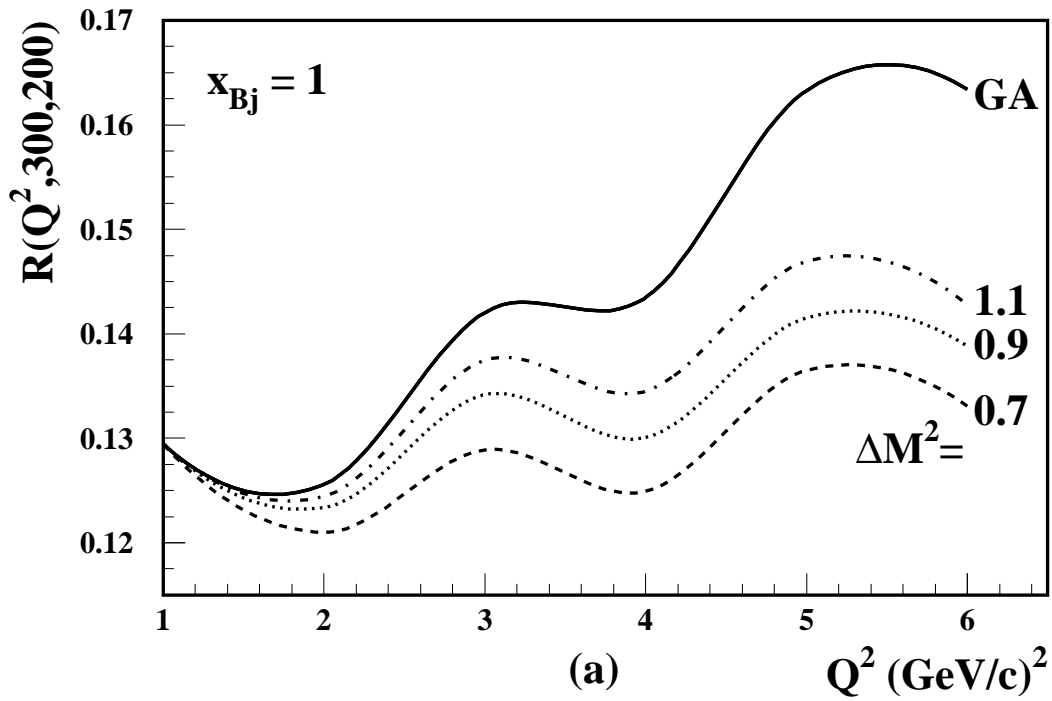
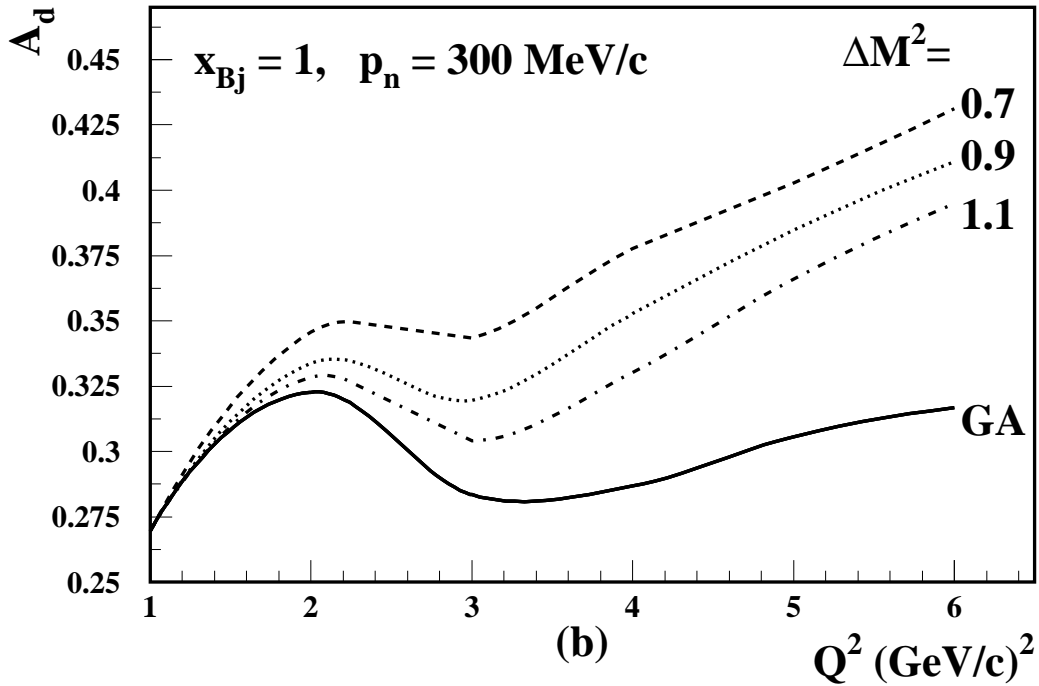
\end{document}